\documentstyle[epsf,12pt,eqsecnum,aps,amssymb]{revtex}

\draft
\begin{document}

\title{Critical adsorption on curved objects}
\author{A. Hanke and S. Dietrich}
\address{Fachbereich Physik, Bergische Universit\"at Wuppertal,\\
D-42097 Wuppertal, Federal Republic of Germany}
\date{\today}
\maketitle
\bigskip
\begin{abstract}
A systematic fieldtheoretic description of critical 
adsorption on curved objects such as spherical 
or rodlike colloidal particles immersed in a fluid near
criticality is presented. The temperature
dependence of the corresponding order parameter profiles
and of the excess adsorption are calculated explicitly.
Critical adsorption on elongated rods is substantially 
more pronounced than on spherical particles.
It turns out that, within the context of critical 
phenomena in confined geometries, critical adsorption on a 
microscopically thin `needle' represents a distinct 
universality class of its own.
Under favorable conditions the results are relevant for 
the flocculation of colloidal particles.
\end{abstract}
\bigskip

\narrowtext

\newpage

\section{Introduction}
\label{secI}

In colloidal suspensions the interaction between the mesoscopic
dissolved particles and the solvent is of basic 
importance \cite{dispersions,colloid}. For example, the solvent 
generates effective interactions between the colloidal particles 
which can even lead to flocculation. The richness of the
physical properties of these systems is mainly based on the 
possibility to tune these effective interactions over wide
ranges of strength and form of the interaction potential.
Traditionally this tuning is accomplished by 
changing the chemical composition of the solvent, e.g., by 
adding salt, polymers, or other components \cite{dispersions}.
Compared with such modifications,
changes of the temperature or pressure typically result only
in minor changes of the effective interactions. This, however, is 
only true as long as the solvent is not thermodynamically close to 
a phase transition of its own. For example, if the solvent consists of 
a binary liquid mixture close to a {\em first-order\/} demixing transition 
into a A-rich and a B-rich liquid phase, even slight changes 
of the temperature or of the partial pressures of the two species
A and B can lead to massive changes of the effective interactions
between dissolved colloid particles induced by the occurrence of
wetting transitions. They lead to wetting 
films of the preferred phase coating the colloidal particles 
\cite{bieker}. These wetting films can snap into bridges if the 
particles come close to each other leading to flocculation 
\cite{beysens,lowen}. For charged colloidal particles such as
silica spheres immersed in the binary liquid mixture of water 
and 2,6-lutidine \cite{beysens} flocculation can 
also be influenced by screening effects generated 
by the adsorbed layers \cite{law12}.

Similarly drastic effects can occur if the solvent is brought 
close to a {\em critical point\/}. The inevitable preference of 
the surfaces of the colloidal particles for one of the two solvent 
species of a binary liquid mixture near its critical demixing point 
or for the liquid phase of a one- or two-component solvent fluid near 
its liquid-vapor critical point results into the presence of effective
surface fields leading to pronounced adsorption profiles of the 
preferred component. This so-called `critical adsorption' becomes 
particularly long-ranged due to the correlation effects induced by 
the critical fluctuations of the order parameter of the solvent. 
In the case of a planar wall critical adsorption has been 
studied in much detail 
\cite{binder,diehl,fisher,cardy90,stapper,ds93,sdl94,floter,slsl97,carpenter}. 
Asymptotically close to the critical point $T_c$
it is characterized by a surface field
which is infinitely large so that the order parameter
profile actually diverges upon approaching the wall up to 
atomic distances $\sigma$.
As compared to a planar surface critical adsorption on a 
spherical particle is expected to exhibit important 
differences in behavior because the confining surface 
has a positive curvature and because a sphere represents 
only a quasi-zero-dimensional defect floating in the 
critical fluid. The interference of critical adsorption 
on neighboring spheres gives rise to the so-called 
critical Casimir forces \cite{krech} which have been 
argued to contribute to the occurrence of flocculation
near $T_c$ \cite{dG81,burkhardt,netz}. A quantitative 
understanding of these phenomena requires 
the knowledge of the critical adsorption profiles near the colloidal 
particles and the resulting effective free energy of interaction 
in the whole vicinity of the critical point, i.e., as
function of both the reduced temperature $t = (T - T_c) / T_c$ and 
the field $h$ conjugate to the order parameter. This ambitious
goal has not yet been accomplished. Instead, the introduction of 
a surface curvature 
has limited the knowledge of the corresponding critical adsorption so 
far to the case of spheres for the particular 
thermodynamic state $(t,h) = (0,0)$ of the solvent
\cite{be85,gnutzmann}. Only recently at least the temperature 
dependence of the critical Casimir force between a sphere and a 
planar container wall has been addressed \cite{prl}. 
Thus the present study of the temperature dependence of
critical adsorption on a single sphere contributes one
step towards reaching the aforementioned general goal.

Apart from spherical particles, also rodlike particles
play an important role \cite{buining}. Rodlike objects are 
provided, e.g., by fibers or colloidal rods \cite{buining}, 
semiflexible polymers with a large persistence length 
such as actin \cite{gittes}, microtubuli \cite{gittes}, 
and carbon nanotubes \cite{nanotube}. 
Moreover the knowledge of the general curvature dependence 
of critical adsorption is also relevant, e.g., for curved 
membranes \cite{nelson,vesicle} dissolved in a
fluid near criticality or for the liquid-vapor interface 
between a binary 
liquid mixture near its critical demixing point and its 
noncritical vapor, which exhibits rippled configurations 
due to the occurrence of capillary waves \cite{nelson}.

Near criticality the relevant length scales of the solvent 
structures are dominated by the diverging bulk correlation length
$\xi_{\pm} = \xi_0^{\pm} |t|^{-\nu}$, where $\nu$ is the standard 
universal bulk critical exponent; $\xi_0^{+}$ and $\xi_0^{-}$
are nonuniversal amplitudes in the one- ($+$) and two-phase region
($-$), respectively, with values typically in the order of a few {\AA}.
In practice the correlation length can span the range between 
$5_{\,}${\AA} - $1_{\,}\mu\text{m}$ depending on $t$.
In the present context this length scale is played off against
the length scale $R$ of the radius of the dissolved particles.    
We note that the available systems can realize both the limit
$R / \xi \gg 1$ as well as the opposite limit $R / \xi \ll 1$. 
In the case of Ludox silica particles $R \approx 12\,\text{nm}$
\cite{kaler} so that the limit $R / \xi \ll 1$ can be easily 
achieved even with the upper limits for $\xi$ set by finite 
experimental resolutions. 
The ratio of the length $l$ and the radius $R$ of rodlike
particles can be quite large, in conjunction with
a small radius such as $R \approx 7\,\text{nm}$ in the case 
of colloidal boehmite rods \cite{buining}. In this work
we consider {\em long\/} rods, i.e., $R, \xi \ll l$, and
neglect effects which may arise due to their finite length $l$. 

\begin{figure}[t]
\begin{picture}(16,10)
\put(-0.7,-1.3){
\setlength{\epsfysize}{11.3cm}
\epsfbox{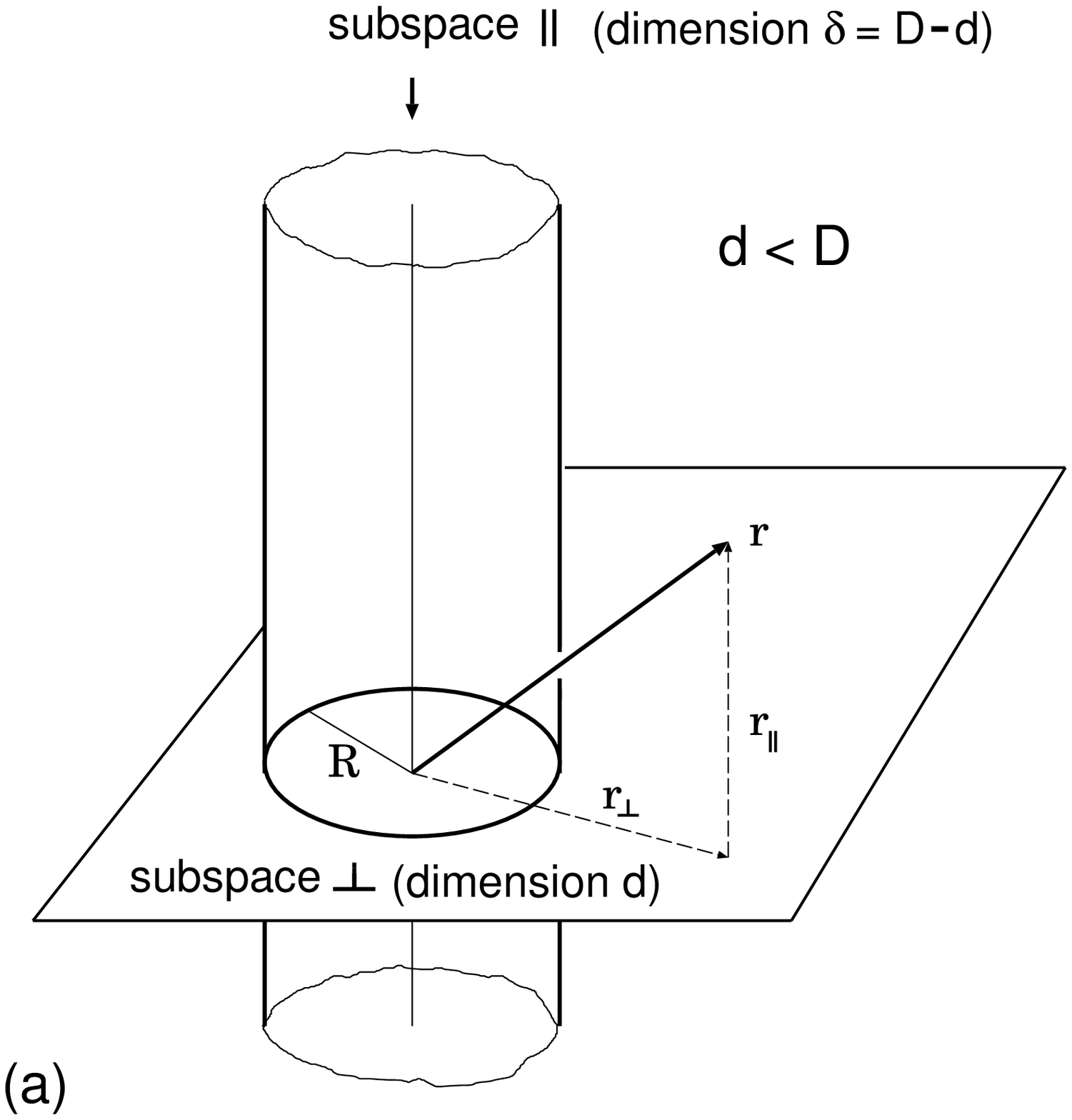}}
\put(8.0,-1.3){
\setlength{\epsfysize}{10cm}
\epsfbox{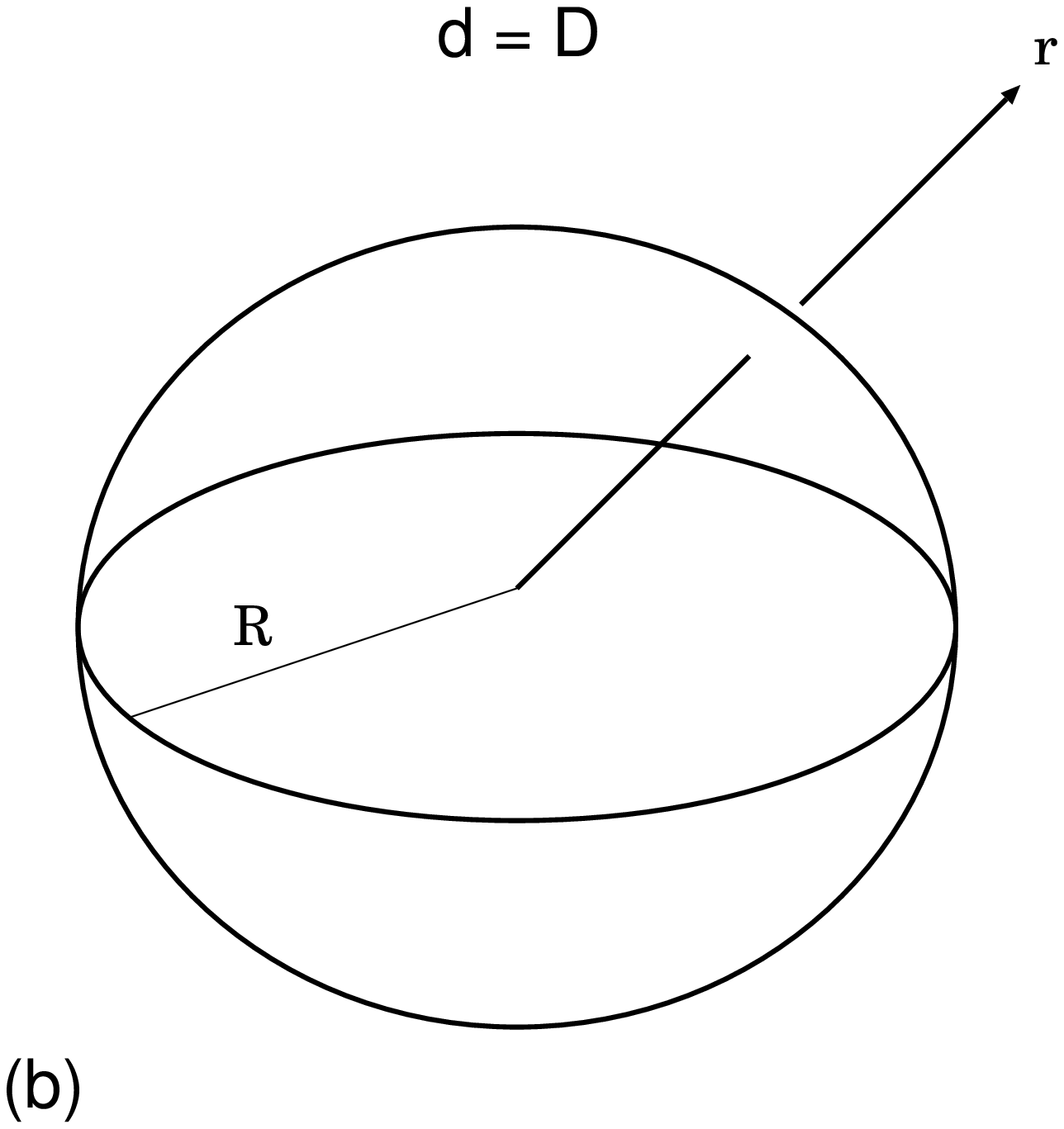}}
\end{picture}
\caption{(a) A generalized cylinder $K$
with $d < D$ (i.e., $\delta = D - d > 0$, see Eq.\,(\ref{I10}))
and (b) a sphere as the special case of $K$ with $d = D$ as examples of 
particles with curved boundaries. $D$ is the spatial dimension. 
The point ${\bf r} = ({\bf r}_{\perp}, {\bf r}_{\parallel}\,)$
at which, e.g., the order parameter is monitored is also shown.}
\label{fig_cyl}
\end{figure}

In the present contribution we investigate systematically 
the temperature dependence of the critical adsorption
on a single spherical or rodlike particle, i.e., the case
$(\protect{t \not= 0}, \protect{h = 0})$. (The generalization 
to the case $h \not= 0$ is straightforward but tedious.)
In order to be able to treat spheres and cylinders
in a unified way within a field-theoretical 
approach and for general spatial dimensions $D$ it is 
helpful to consider the particle shape of a 
{\em generalized cylinder\/} $K$ \cite{context}
with an infinitely extended `axis' of dimension $\delta$.
The `axis' can be the axis of an ordinary infinitely 
elongated cylinder ($\delta = 1$), or the midplane of 
a slab ($\delta = D - 1$), or the center of a sphere 
($\delta = 0$). For general integer $D$ and $\delta$ 
the explicit form of $K$ is
\begin{equation} \label{I10} 
K \, = \, \Big\{ {\bf r} = ({\bf r}_{\perp},
{\bf r}_{\parallel} \,) \in {\mathbb R}^{D - \delta}
\times {\mathbb R}^{\delta}; \, \,  | {\bf r}_{\perp} | \le R \Big\}
\end{equation}
with ${\bf r}_{\perp}$ and ${\bf r}_{\parallel}$ perpendicular and
parallel to the axis, respectively (see Fig.\,\ref{fig_cyl}).
Note that ${\bf r}_{\perp}$ is a $d$-dimensional vector with 
\begin{equation} \label{I15} 
d = D - \delta \, \, .
\end{equation}
The radius $R$ of the generalized cylinder $K$ is the radius in 
the cases of an ordinary cylinder or a sphere and it is half of
the thickness in the case of a slab. For the slab the geometry 
reduces to the much studied case of (two decoupled) half spaces. 
The generalization of $D$ to values 
different from three is introduced for technical reasons 
because $D_{uc} = 4$ marks the upper critical dimension 
for the relevance of fluctuations of the order parameter leading
to a behavior different from that obtained from mean-field theory
valid for $D = 4$.

In Sec.\,\ref{sec_op} we discuss the general scaling properties
of the local order parameter profiles 
for critical adsorption on spheres and cylinders,
in particular the behavior close to the particle surfaces and for 
small particle radii, respectively. In Sec.\,\ref{sec_excess}
we consider the corresponding properties of the excess 
adsorption. In Sec.\,\ref{sec_mf} we present explicit results
both for the order parameter profiles and for the excess adsorption
in mean-field approximation. Section \ref{summary} contains
our conclusions. In Appendix \ref{app_ex} we discuss the two-point 
correlation function near a microscopically thin `needle' at
criticality. In Appendix \ref{app_neu} we determine a universal 
amplitude and a universal scaling function as needed
in Secs.\,\ref{sec_op} and \ref{sec_excess}, respectively.
In Appendix \ref{appB}, finally, we consider the general curvature
dependence of the excess adsorption.

%
\unitlength1cm
%

\section{Order parameter profiles}
\label{sec_op}

\subsection{General scaling properties}
\label{subsec_allgemein}

Asymptotically, i.e., in case of an infinitely large 
surface field, critical adsorption on the surface of a sphere 
or cylinder with radius $R$ is characterized by an order 
parameter profile 
$\langle \Phi({\bf r}) \rangle_{t}$ which close to $T_c$ 
and for $h = 0$ takes the scaling form
\begin{equation} \label{A10}
\langle \Phi({\bf r}) \rangle_{t} = a |t|^{\beta} 
P_{\pm}( s / \xi_{\pm}, R / \xi_{\pm} )
\end{equation}
for radial distances $s = r_{\perp} - R \gtrsim \sigma$ 
from the surface 
larger than a typical microscopic length $\sigma$. Here 
$\langle \, \, \, \rangle_{t}$ denotes the thermal average in the 
presence of a sphere or a cylinder
and $\beta$ is the standard universal bulk critical exponent. 
The scaling functions $P_{\pm}$ depend on two scaling variables:
\begin{equation} \label{A20}
x_{\pm} \, = \, s / \xi_{\pm} \, \, ,
\quad y_{\pm} \, = \, R / \xi_{\pm} \, \, .
\end{equation}
The scaling functions $P_{\pm}$ are {\em universal\/} while the 
nonuniversal {\em bulk} amplitudes $a$ and $\xi_{0}^{\pm}$ are
determined by the value 
$\langle \Phi \rangle_{t,\,b} = a |t|^{\beta}$ for $t < 0$, 
$t \to 0^{-}$ of the order parameter in the unbounded bulk and by the 
amplitudes of the correlation length, respectively. (In this work
we are always concerned with the Ising universality class.) 
We note that the form of $P_{\pm}$ depends on the definition used
for the correlation length $\xi_{\pm}$. For definiteness we adopt
$\xi_{\pm}$ as being the {\em true\/} correlation length 
fixed by the exponential decay of the two-point correlation function 
in real space \cite{definition}. Bearing this in mind one finds that
$P_{+, \, b} \equiv P_{+}(\infty, y_{+} ) = 0$, 
$P_{-, \, b} \equiv P_{-}(\infty, y_{-} ) = 1$, and 
$P_{\pm}(x_{\pm} \to \infty, y_{\pm}) - P_{\pm, \, b} 
\sim \exp(- x_{\pm})$ where the prefactor in front of the exponential
may by an algebraic function of $x_{\pm}$
(see, c.f., Sec.\,\ref{sec_mf}).

On the other hand, in the limit $y_{\pm} \to \infty$ with
$x_{\pm}$ fixed the scaling functions 
$P_{\pm}(x_{\pm}, y_{\pm})$ reduce to
\begin{equation} \label{A25}
P_{\pm}(x_{\pm}) \, \equiv \, P_{\pm}(x_{\pm}, \infty) \, \, ,
\qquad \text{half-space} \, \, ,
\end{equation}
corresponding to the half-space bounded by a planar surface. 
In recent years the scaling functions $P_{\pm}(x_{\pm})$ 
for the half-space have been intensively studied theoretically 
\cite{fisher,ds93,sdl94,floter} and compared with 
experiments \cite{fisher,floter,slsl97,carpenter}. 
For later reference we quote some of their properties. 
First, we note the asymptotic behavior \cite{ds93}
\begin{eqnarray}
& & P_{\pm}(x_{\pm} \to 0) \, \to \, c_{\pm} \, 
x_{\pm}^{\, - \beta/\nu} \, \label{A40}\\[1mm]
& & \, \, \times \, 
\Big[ 1 \, + \, \bar{a}_{\pm} \, x_{\pm}^{\,1/\nu} \, + \, 
\bar{a}_{\pm}^{\,'} \, x_{\pm}^{\,2/\nu} \, + \, 
\bar{b}_{\pm} \, x_{\pm\,}^{\,D} \, + \, \ldots \, \Big] \nonumber
\end{eqnarray}
with universal amplitudes $c_{\pm}$, $\bar{a}_{\pm}$, 
$\bar{a}_{\pm}^{\,'}$, and $\bar{b}_{\pm}$\cite{bar} which 
depend, however, on the definition of the correlation length 
\cite{definition}. The ellipses stand for contributions which 
vanish more rapidly than $x_{\pm}^{\,D}$.\\[1mm]

%
\unitlength1cm
\begin{figure}[t]
\caption{(a) Universal scaling function $P_{+}(x_{+})$
for critical adsorption on a planar surface in
spatial dimensions $D = 4$, $3$, and $2$. The estimates for
$P_{+}(x_{+})$ in $D = 3$ labelled `MC' and `RG' are 
obtained by Monte Carlo simulations and field-theoretical
renormalization-group calculations. We display the 
corresponding data presented in Ref.\,[15]. The inset shows 
a scaled version of $P_{+}(x_{+})$ which reflects, in particular,
the asymptotic behavior of $P_{+}(x_{+} \to 0)$. The open symbols 
indicate the corresponding values of $c_{+}$ (see Eq.\,(\ref{A40}) 
and Table\,\ref{tab_ca}). (b) Same representation as in
(a) for the universal scaling function $P_{-}(x_{-})$.
The inset shows a scaled version of $P_{-}(x_{-})$ which
contains the function $U(x_{-})$ defined in Eq.\,(\ref{A61})
in order to facilitate a similar representation as in the 
inset of (a).}
\begin{picture}(16,20)
\put(-0.5,12.3){
\setlength{\epsfysize}{10cm}
\epsfbox{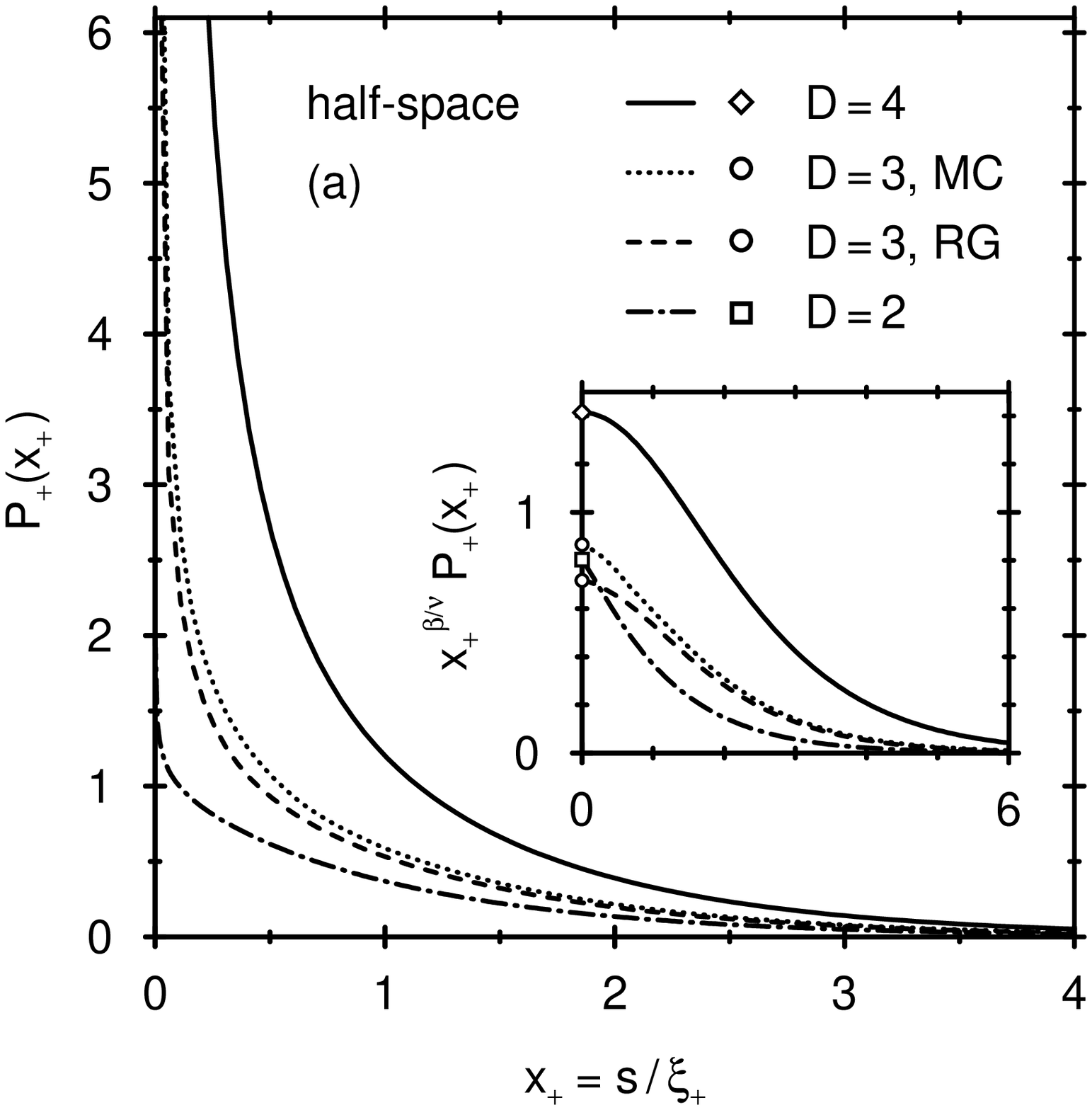}}
\put(-0.5,0.1){
\setlength{\epsfysize}{10cm}
\epsfbox{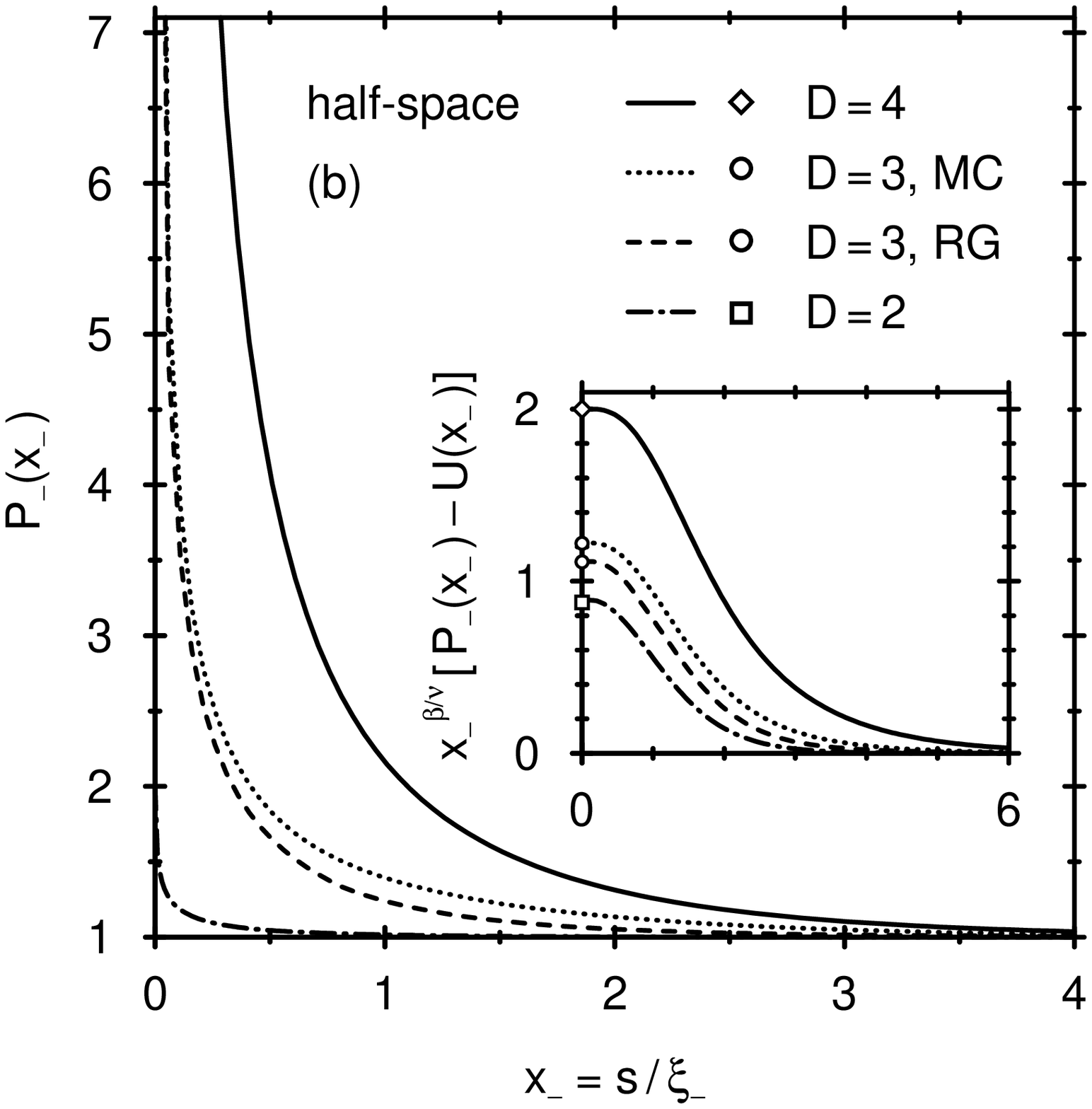}}
\end{picture}
\label{fig_hs}
\end{figure}
%
%
\newpage
\noindent
The exponent $\beta / \nu$
is the bulk `scaling dimension' \cite{zinn} of the order parameter
$\Phi$. For convenience of the reader we quote their numerical values 
\cite{zinn,fisher}:
\begin{mathletters}
\label{A45}
\begin{equation} \label{A45a}
\beta(D=4) = 1/2      \, \, , \quad
\beta(3) \simeq 0.328 \, \, , \quad
\beta(2) = 1/8        \, \, ,
\end{equation}
\begin{equation} \label{A45b}
\nu(D=4) = 1/2      \, \, , \quad
\nu(3) \simeq 0.632 \, \, , \quad
\nu(2) = 1          \, \, ,
\end{equation}
so that 
\begin{equation} \label{A45c}
(\beta/\nu) \, (D=4) = 1        \, \, , \quad
(\beta/\nu) \, (3) \simeq 0.519 \, \, , \quad
(\beta/\nu) \, (2) = 1/8        \, \, .
\end{equation}
The leading power law in Eq.\,(\ref{A40}) is a 
consequence of the fact that due to the presence 
of the symmetry breaking surface field the limit 
$t \to 0$ must lead to a nonvanishing order parameter 
profile at $t = 0$ so that
\end{mathletters}
\begin{equation} \label{A50}
\langle \Phi({\bf r}) 
\rangle_{\text{hs}, \, t \, = \, 0} \, = \,
a \, c_{\pm} \, (s / \xi_{0}^{\pm})^{- \beta/\nu} \, \, .
\end{equation}
(The subscript `hs' stands for half-space).
Thus the amplitude of the power law $s^{- \beta/\nu}$ in  
Eq.\,(\ref{A50}) consists of the combination 
$a_{\,} (\xi_{0}^{\pm})^{\beta/\nu}$ of
nonuniversal bulk amplitudes and of the universal surface 
amplitude $c_{\pm}$. Because the lhs of
Eq.\,(\ref{A50}) does not depend on how the limit $t = 0$ is
approached, the universal amplitudes 
$c_{+}$ and $c_{-}$ are related via the universal 
bulk ratio $\xi_{0}^{+} / \xi_{0}^{-}$:
\begin{equation} \label{A55} 
\frac{c_{-}}{c_{+}} \, = \, 
\Bigg( \frac{\xi_{0}^{+}}{\xi_{0}^{-}} \Bigg)^{\beta/\nu} \, \, .
\end{equation}
The terms proportional to $\bar{a}_{\pm}$ and $\bar{a}_{\pm}^{\,'}$ in 
Eq.\,(\ref{A40}) correspond to {\em regular\/} contributions 
for $t \to 0$ of the order parameter since
\begin{equation} \label{A60}
\langle \Phi({\bf r}) \rangle_{\text{hs}, \, t} \, \to \, 
\langle \Phi({\bf r}) \rangle_{\text{hs}, \, t \, = \, 0 \,}
\left[1 + A \, t + A'\,t^2 + B_{\pm} |t|^{2 - \alpha} \, + \, \ldots
\, \right] \, \, , \quad t \to 0 \, \, ,
\end{equation}
where
$A = \pm \, \bar{a}_{\pm} (s / \xi_{0}^{\pm})^{1/\nu}$ and
$A' = \bar{a}_{\pm}^{\,'} (s / \xi_{0}^{\pm})^{2/\nu}$ are 
{\em independent\/} of the sign of $t$ \cite{ds93}. The term 
proportional to $\bar{b}_{\pm}$ in Eq.\,(\ref{A40}) corresponds 
to the first singular contribution \cite{tensor}
for $t \to 0$ of the order parameter
leading to the term $B_{\pm} |t|^{2 - \alpha}$ in Eq.\,(\ref{A60})
with the bulk exponent $\alpha = 2 - \nu D$ and
$B_{\pm} = \bar{b}_{\pm} (s / \xi_{0}^{\pm})^D$.

Figure \ref{fig_hs} summarizes theoretical results for 
$P_{\pm}(x_{\pm})$ for the spatial dimensions 
$D = 4$, $3$, and $2$. In $D = 4$ the functions $P_{\pm}$
are given by the following mean-field expressions: 
\begin{equation} \label{A30}
P_{+}(x_+) \, = \, 
\frac{\sqrt{2}}{\sinh({x_+})} \, \, , 
\quad P_{-}(x_-) \, = \, 
\coth\left( \frac{x_-}{2} \right) \, \, ,
\qquad D = 4 \, \, .
\end{equation}
The results for $D = 3$ represent
recent Monte Carlo (MC) simulations \cite{sdl94,slsl97} and 
field-theoretical renormalization group (RG) calculations
\cite{ds93,slsl97}. For $D = 2$ exact results are available 
from the semi-infinite two-dimensional Ising model \cite{bariev}. 
Note that $P_{\pm}(x_{\pm})$ for any fixed 
value of $x_{\pm}$ {\em decreases\/} with decreasing $D$ 
\cite{coth}. This reflects the general trend that critical 
fluctuations, which reduce the mean value of the order 
parameter, are more pronounced in lower spatial dimensions. 

The curves in the inset of Fig.\,\ref{fig_hs} are 
scaled so that they demonstrate the leading behavior of 
$P_{\pm}(x_{\pm} \to 0)$. (A similar representation will be
used in Sec.\,\ref{sec_mf}.)
In Table \ref{tab_ca} we quote the corresponding numerical values 
of the surface amplitudes $c_{\pm}$ and $\bar{a}_{\pm}$
according to Eq.\,(\ref{A40}).
In order to achieve a presentation in the inset of 
Fig.\,\ref{fig_hs}\,(b) which reflects both the leading behavior of 
$P_{-}(x_{-} \to 0)$ and the exponential decay  
$P_{-}(x_{-} \to \infty) - 1 \sim \exp(-x_{-})$
we introduce the function 
\begin{equation} \label{A61}
U(x_{-}) \, = \, \tanh\Big( \frac{x_{-}^{\,2}}{x_{-} + 1} \Big) \, \, .
\end{equation}
The curves in the inset of Fig.\,\ref{fig_hs}\,(b) comply with 
the above condition since $U(x_{-} \to 0) \sim x_{-}^{\,2}$ and 
$U(x_{-} \to \infty) - 1 \sim \exp(- 2 x_{-})$ so that in both limits
the leading asymptotic behaviors of $P_{-}(x_{-})$ are not 
changed.\\[1mm]


\begin{table}[tbp]
\medskip
\caption{Numerical values of the universal amplitudes $c_{\pm}$ 
and $\bar{a}_{\pm}$ (see Eq.\,(\ref{A40})).}
\label{tab_ca}
\bigskip
\begin{tabular}{lcccccc} 
$D$ & $c_{+}$ & $c_{-}$ & $\bar{a}_{+}$ & $\bar{a}_{-}$ \\ \hline
$4$     & $\sqrt{2} \simeq 1.414$ & $2$     
&   $-\frac{1}{6} \simeq -0.167$ & $\frac{1}{12} \simeq 0.083$ \\
$3$, MC\tablenote{Reference \cite{slsl97}} & $0.866$ & $1.22$  \\
$3$, RG\tablenote{Reference \cite{slsl97}; for $\bar{a}_{\pm}$
we use in addition Eq.\,(48) in Ref.\,\cite{ds93} for
$\varepsilon = 1$}
& $0.717$ & $1.113$ & $-0.389$ & $0.129$ \\ 
$3$, interpolation\tablenote{Reference \cite{floter}; obtained from
interpolating between the results in $D = 4 - \varepsilon$ and $D = 2$}
& $0.94 \pm 0.05$ & $1.24 \pm 0.05$ \\ 
$2$ & $0.803$  & $0.876$ & $- \frac{1}{2} = - 0.5$ 
& $\frac{1}{4} = 0.25$ \\ 
\end{tabular}
\end{table}


\newpage

In the remaining part of this section we
discuss the new features of $P_{\pm}(x_{\pm}, y_{\pm})$
which arise for $y_{\pm} < \infty$. First, we consider the short
distance behavior for $s \to 0$, which corresponds to the limit 
$x_{\pm} \to 0$ with $y_{\pm}$ fixed. Then, we consider cases
in which the radius $R$ is small compared with $\xi_{\pm}$ 
as well as the distance between the sphere or the cylinder and the point 
for which the order parameter is monitored. This corresponds to 
the limit $y_{\pm} \to 0$ with $x_{\pm}$ fixed. It turns out that
in this limit the behaviors for a sphere and a cylinder in $D = 3$ 
are {\em qualitatively\/} different.

\subsection{Short distance expansion}
\label{subsec_short}

The same reasoning leading to Eq.\,(\ref{A50}) in case of the 
half-space yields for the generalized cylinder
\begin{equation} \label{A62}
\langle \Phi({\bf r}) \rangle_{t \, = \, 0} \, = \, 
a \, {\cal C}_{\pm}(\gamma) \, 
(s / \xi_{0}^{\pm})^{- \beta/\nu} \, \, , \quad \gamma = s / R \, \, ,
\end{equation}
i.e., the universal amplitude $c_{\pm}$ in Eq.\,(\ref{A50})
is generalized to the universal amplitude {\em function\/} 
${\cal C}_{\pm}(\gamma)$ of the scaling variable $\gamma$.
This function appears in the asymptotic behavior
\begin{equation} \label{A65}
P_{\pm}(x_{\pm}, y_{\pm} = x_{\pm} / \gamma) \, \to \,
{\cal C}_{\pm}(\gamma) \, \, x_{\pm}^{- \beta / \nu} \, \, ,
\quad x_{\pm} \to 0 \, , \, \, \gamma \, \, \mbox{fixed} \, \, ,
\end{equation}
which underscores that ${\cal C}_{\pm}(\gamma)$ is universal
but depends on the definition of the correlation length
\cite{definition} (compare Eq.\,(\ref{A40})) and on the geometry.
Since the limit $\gamma \to 0$ must reproduce the behavior 
for the half-space one has ${\cal C}_{\pm}(0) = c_{\pm}$. 
According to Eq.\,(\ref{A55}) the functions ${\cal C}_{+}(\gamma)$ 
and ${\cal C}_{-}(\gamma)$ are proportional to each other with 
\begin{equation} \label{A70}
\frac{{\cal C}_{-}(\gamma)}{{\cal C}_{+}(\gamma)} \, = \, 
\frac{c_{-}}{c_{+}} \, = \, 
\Bigg( \frac{\xi_{0}^{+}}{\xi_{0}^{-}} \Bigg)^{\beta/\nu} \, \, .
\end{equation}
Therefore it is sufficient to study only one of these functions, 
say, ${\cal C}_{+}(\gamma)$, which according to Eqs.\,(\ref{A62})
and (\ref{A50}) can be written as
\begin{equation} \label{A80}
{\cal C}_{+}(\gamma) \, = \, c_{+} \, 
\langle \Phi({\bf r}) \rangle_{t \, = \, 0} \, / \,  
\langle \Phi({\bf r}) 
\rangle_{\text{hs}, \, t \, = \, 0} \, \, .
\end{equation}
In the case of a sphere, i.e., $d = D$, the universal scaling
function ${\cal C}_{+}(\gamma)$ is known exactly for any
spatial dimension $D$ of interest by means of a finite 
conformal mapping from the half-space \cite{be85}. 
It takes the simple analytic form
\begin{equation} \label{A85}
{\cal C}_{+}(\gamma) \, = \, c_+ \, 
\Big( 1 + \frac{\gamma}{2} \Big)^{- \beta / \nu} \, \, ,
\quad d = D \, \, ,
\end{equation}
and depends on $D$ only via the corresponding 
values of $c_+$ and $\beta / \nu$.
For large distances from the sphere, i.e. $\gamma \gg 1$, the 
function ${\cal C}_{+}(\gamma)$ in Eq.\,(\ref{A85}) decays as 
$\gamma^{-\beta / \nu}$ (compare the following subsection).
In the opposite limit $\gamma \to 0$, which is of interest
in the present subsection, one has
\begin{equation} \label{A90}
\frac{{\cal C}_{+}(\gamma \to 0)}{c_{+}} \, = \,
1 \, - \, \frac{\beta / \nu}{2} \, \gamma \, + \,
\frac{\beta / \nu \, (\beta / \nu + 1)}{8} \, \gamma^2 
\, + \, {\cal O}(\gamma^3) \, \, , \quad d = D \, \, .
\end{equation}

We note, first, that an expansion 
such as in Eq.\,(\ref{A90}) is expected to hold not only near the surface 
of a sphere but also near a smoothly curved surface of 
{\em more general shape}. According to differential geometry 
up to second order in curvature
the short distance expansion near a ($D - 1$)-dimensional surface 
of arbitrary shape involves only  
the geometric invariants $K_m$, $K_m^{\,2}$, and $K_{G}$ with 
\cite{david,geometry}
\begin{mathletters}
\label{A100}
\begin{equation} \label{A100a}
K_m \, = \, \frac{1}{2} \, \sum_{i=1}^{D-1} \, \frac{1}{R_i}
\end{equation}
and 
\begin{equation} \label{A100b}
K_G \, = \, \sum_{pairs \atop i<j}^{D-1} \, \frac{1}{R_i R_j}
\end{equation}
where $R_i$ are the $D - 1$ principal local radii of curvature.
In $D = 3$ Eq.\,(\ref{A100}) yields the familiar expressions 
for the local mean curvature and the local Gaussian curvature.
Second, we note that Eq.\,(\ref{A90}) reflects a property of the 
fluctuating order parameter field $\Phi({\bf r})$ (or `operator') 
in the outer space of the sphere itself. In this spirit 
also near a surface of more general shape the short 
distance expansion can be formulated in `operator form':
\end{mathletters}
\begin{eqnarray}
& & \Phi({\bf r}) / 
\langle \Phi({\bf r}) \rangle_{\text{hs}, \, t \, = \, 0}
\label{A110}\\[1mm]
& & \, = \, [{\bf 1}] \, \Big\{1 \, + \,
\kappa_1 \, K_{m} \, s \, + \, 
\kappa_2 \, K_m^{\,2} \, s^2 
\, + \, \kappa_G \, K_G \, s^2 \, + \, {\cal O}(s^3) \Big\}
\, + \, {\cal O}(s^D) \nonumber \, \, .
\end{eqnarray}
Here $s$ is the distance of ${\bf r}$ to the nearest point 
of the surface. The terms on the rhs of Eq.\,(\ref{A110})
should be interpreted 
as operators which are located {\em at\/} this point of the 
surface, i.e., as {\em surface operators\/}. On the rhs of
Eq.\,(\ref{A110}) only such terms are shown explicitly 
which are proportional to the unity operator $[{\bf 1}]$.   
The corrections emerge from curvatures of higher order 
and surface operators of different types which 
are expected to scale with 
powers of $s$ at least of the order $s^D$
(compare Eq.\,(\ref{A40}) for the half-space 
and Ref.\,\cite{tensor}). Thus for $D > 2$ 
the terms displayed in Eq.\,(\ref{A110}) represent the leading 
contributions of the short distance expansion of $\Phi({\bf r})$.
The dimensionless coefficients $\kappa_1$,
$\kappa_2$, and $\kappa_G$ depend on $D$ but not on the shape 
of the boundary surface. Comparison of Eq.\,(\ref{A110}) with 
Eq.\,(\ref{A90}) for the sphere, or direct calculation near a 
surface of arbitrary shape \cite{cardy90}, yields
\begin{equation} \label{A120}
\kappa_1 \, = \, - \, \frac{\beta / \nu}{D - 1} \, \, .
\end{equation}
In $D = 3$
the coefficients $\kappa_2$ and $\kappa_G$ in Eq.\,(\ref{A110})
cannot be determined by comparison with Eq.\,(\ref{A90}) 
for the sphere alone. To this end one would need, in addition, 
the knowledge of the expansion for at least one curved surface 
of different shape (e.g., the surface of a cylinder). However, 
since in $D = 3$ the sphere has the property $K_G = K_{m\,}^{\,2}$
the comparison of Eq.\,(\ref{A110}) with Eq.\,(\ref{A90})
determines at least the sum
\begin{equation} \label{A130}
\kappa_2 \, + \, \kappa_G \, = \, 
\frac{\beta / \nu \, (\beta / \nu + 1)}{8} \, \, ,
\quad D = 3 \, \, .
\end{equation}
In Sec.\,\ref{sec_mf} we confirm Eq.\,(\ref{A110}) and determine 
$\kappa_1$, $\kappa_2$, and $\kappa_G$ for $D = 4$. 

The expansion (\ref{A110}) holds 
upon inserting it into thermal averages
if the distance $s$ to the curved surface 
$-$ albeit being large on the microscopic scale $-$
is much smaller then other characteristic length scales
such as the correlation length or the distances to the 
remaining operators in correlation functions. For certain 
thermal averages such as the profile 
$\langle \Phi({\bf r}) \rangle_{t}$
additional {\em regular\/} terms can occur on the rhs
(compare Eq.\,(\ref{A60}) for the half-space).
For spheres and cylinders, in particular, 
the expansion (\ref{A110}) determines 
the leading contributions to the scaling 
functions $P_{\pm}(x_{\pm}, y_{\pm})$
generated by the surface curvature.
For the surface of a generalized cylinder the 
curvatures in Eq.\,(\ref{A100}) are given by
\begin{mathletters}
\label{A140}
\begin{equation} \label{A140a}
K_m \, = \, \frac{d-1}{2} \, \frac{1}{R}
\end{equation}
and 
\begin{equation} \label{A140b}
K_G \, = \, \frac{(d-1)(d-2)}{2} \, \frac{1}{R^2} \, \, .
\end{equation}
By inserting these expressions into Eq.\,(\ref{A110}) and 
using Eq.\,(\ref{A10}) one obtains
\end{mathletters}
\begin{eqnarray}
& & P_{\pm}(x_{\pm} \to 0, y_{\pm}) \Big|_{\text{sde}} \, \to \,
c_{\pm} \, x_{\pm}^{\, - \beta / \nu} \, 
\Big\{ 1 \, + \, \kappa_1 \, \frac{d-1}{2} \, \gamma
\label{A148}\\[2mm]
& & \, + \, \Big[ \kappa_2 \, \frac{(d-1)^2}{4} \, + \,
\kappa_G \, \frac{(d-1)(d-2)}{2} \Big] \, \gamma^2 \,
\, + \, {\cal O}(\gamma^3) \, \Big\} \nonumber
\end{eqnarray}
for $x_{\pm} \to 0$ with $y_{\pm}$ fixed so that 
$\gamma = x_{\pm} / y_{\pm} \to 0$ (the subscript `sde'
refers to the short distance operator expansion in 
Eq.\,(\ref{A110})). 

Note that Eq.\,(\ref{A148}) does not contain the above mentioned 
regular contributions. Upon employing Eq.\,(\ref{A40}) for 
the half-space, however, one can obtain the asymptotic expansion 
of $P_{\pm}(x_{\pm} \to 0, y_{\pm})$ including the leading regular 
term. Assuming that for $t \gtrless 0$ 
the profile $\langle \Phi({\bf r}) \rangle_{t}$
is still analytic in $1/R$ one finds
\begin{eqnarray}
& & P_{\pm}(x_{\pm} \to 0, y_{\pm}) \, \to \,
c_{\pm} \, x_{\pm}^{\, - \beta / \nu} \, 
\Big\{ 1 \, + \, \kappa_1 \, \frac{d-1}{2} \, \gamma
\label{A150}\\[2mm]
& & \, + \, \Big[ \kappa_2 \, \frac{(d-1)^2}{4} \, + \,
\kappa_G \, \frac{(d-1)(d-2)}{2} \Big] \gamma^2 \,
\, + \, \bar{a}_{\pm\,} x_{\pm}^{\,1/\nu} \, + \, \ldots \Big\} 
\nonumber
\end{eqnarray}
for $x_{\pm} \to 0$ with $y_{\pm}$ fixed so that
$\gamma = x_{\pm} / y_{\pm} \to 0$. 
The ellipses stand for contributions 
which vanish more rapidly than $x_{\pm}^{\,2}$ if $D > 2$ \cite{rapid}.
The universal amplitudes 
$c_{\pm}$ and $\bar{a}_{\pm}$ are the same as in Eq.\,(\ref{A40}) 
(see Table\,\ref{tab_ca}) and $\kappa_1$, $\kappa_2$, and $\kappa_G$
are from Eqs.\,(\ref{A120}) and (\ref{A130}). In Sec.\,\ref{sec_mf} we
shall confirm Eq.\,(\ref{A150}) explicitly for $D = 4$.


\subsection{Spheres and cylinders with small radii}
\label{subsec_sre}

In this subsection we consider spheres and cylinders 
whose radii are 
much smaller than other characteristic lengths such as 
the correlation length or the distance between the 
particle and the point at which the order 
parameter is monitored. In these limiting cases the effect 
of the particle upon the fluctuating order parameter 
distribution can be represented by a $\delta$-function 
potential located at the center of the particle which 
enhances the value of the order parameter. 
It is instructive to 
consider this expansion for the generalized cylinder $K$
for which this $\delta$-function 
potential is smeared out over its axis, i.e., the
Boltzmann weight $\exp(- \delta {\cal H}_K)$ of $K$,
where $\delta {\cal H}_K$ is the
difference of the Hamiltonian describing the system with
and without the presence of $K$ (whose axis includes the
origin), can be systematically expanded in a series with
increasing powers of $R$ \cite{burkhardt,prl}, i.e.,
\begin{equation} \label{A170}
\exp(- \delta {\cal H}_K) \, \propto \, 
1 \, + \, {\cal E}_{d,D} \, R^{\beta / \nu - D + d} \, 
w_K \, + \, \ldots
\end{equation}
where ${\cal E}_{d,D}$ is an amplitude and
\begin{equation} \label{A180}
w_K \, = \, \left\{ \begin{array}{l@{\quad}l}
\int\limits_{{\mathbb R}^{\delta}}
d^{\, \delta} r_{\parallel} \, \,
\Phi({\bf r}_{\perp}=0, {\bf r}_{\parallel})
\, \, ,    & d < D \, \, , \\[5mm]
\, \, \Phi(0) \, \, ,  & d = D \, \, .
\end{array} \right.
\end{equation}
Here only the leading nontrivial contribution for $R \to 0$
is shown explicitly and the ellipses stand for contributions 
which vanish more rapidly for $R \to 0$. For the case that 
$K$ is a {\em sphere\/} \cite{burkhardt},
i.e., $d = D$, the amplitude ${\cal E}_{D,D}$ is equal to 
the ratio $A_{\,\uparrow}^{\Phi} / B_{\Phi\,}$, where
$A_{\,\uparrow}^{\Phi}$ and $B_{\Phi}$ are  
amplitudes of the half-space profile 
\begin{equation} \label{A190}
\langle \Phi({\bf r}) \rangle_{\text{hs}, \, t \, = \, 0} \, = \, 
A_{\,\uparrow}^{\Phi\,} (2 s)^{\, - \beta / \nu}
\end{equation}
{\em at\/} the critical point of the fluid for the boundary 
condition $\uparrow$ corresponding to the critical adsorption
fixed point and of the bulk two-point correlation function
\begin{equation} \label{A200}
\langle \Phi({\bf r}) \Phi(0) \rangle_{b, \, t \, = \, 0} \, = \,
B_{\Phi\,} r^{- \, 2 \beta / \nu}
\end{equation}
at criticality, respectively. The ratio 
$(A_{\,\uparrow}^{\Phi})^2 / B_{\Phi}$ is universal \cite{prl}.
The comparison of Eq.\,(\ref{A190}) with Eq.\,(\ref{A50})
yields the relation
\begin{equation} \label{A210}
A_{\,\uparrow}^{\Phi} \, = \,
a \, (2 \xi_{0}^{\pm})^{\beta/\nu} \, c_{\pm} 
\end{equation}
between the nonuniversal amplitudes $A_{\,\uparrow}^{\Phi}$, 
$a$, $\xi_{0}^{\pm}$ and the universal amplitude $c_{\pm}$.

\begin{figure}[t]
\begin{picture}(16,10)
\put(0,0.5){
\setlength{\epsfysize}{9.5cm}
\epsfbox{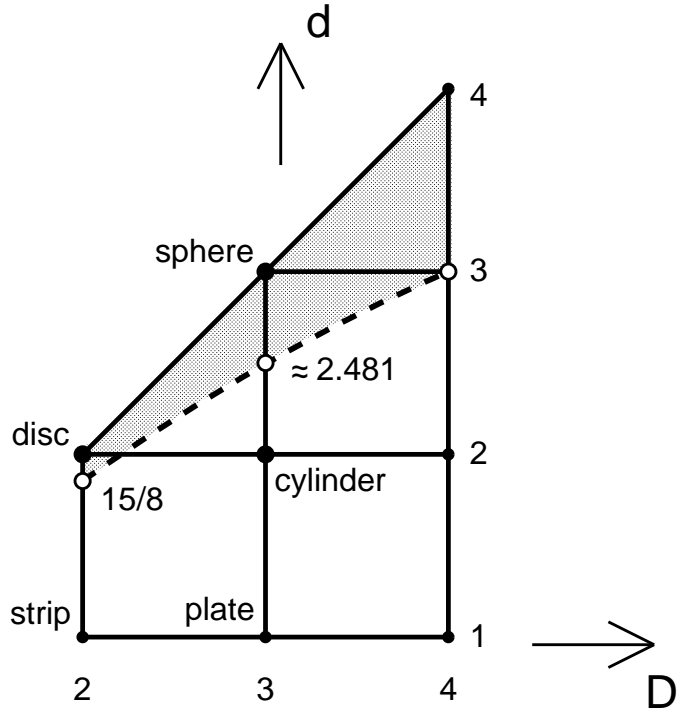}}
\end{picture}
\caption{Diagram of generalized cylinders $K$ which behave
$-$ in the renormalization group sense $-$ as relevant or
irrelevant perturbations of a fluid near criticality. 
The parameter $d \le D$ characterizes the shape of 
$K$ and $D$ is the space dimension (see Eq.\,(\ref{I10})).
The point $(d,D) = (2,2)$ corresponds to a disc in $D=2$ and 
the points $(3,3)$ and $(2,3)$ to a sphere and an 
infinitely elongated cylinder in $D=3$, respectively. 
The line with $D = 4$ and arbitrary $d$ represents
the upper critical dimension for which the mean-field 
results for the adsorption profiles are exact (see 
Sec.\,\ref{sec_mf}). The open circles indicate points $(d,D)$
for which $d = D - (\beta / \nu) (D)$ 
within the Ising universality class.
These points are connected by the dashed line so 
that within the shaded region {\em above\/} it
the small radius expansion (\ref{A170}) is valid 
(see Eq.\,(\ref{A220})) and 
$K$ represents an irrelevant perturbation. 
Points $(d,D)$ below the broken line, such as the cylinder in 
$D = 3$, are characterized by the fact that the 
order parameter at large distances from $K$
deviates from its bulk value even if the radius $R$
is microscopically small, which means that
$K$ represents a relevant perturbation.}
\label{diagram}
\end{figure}

It is crucial to observe that Eq.\,(\ref{A170}) is only valid
if the exponent of $R$ is {\em positive}, i.e.,
\begin{equation} \label{A220}
\beta / \nu \, - \, D \, + \, d \, > \, 0 \, \, .
\end{equation}
Figure \ref{diagram} shows as a dashed line 
$d = D - (\beta / \nu) (D)$ \cite{shape}
in the $(d,D)$-plane which separates generalized cylinders $K$ 
which are {\em relevant\/} perturbations for the fluctuating
order parameter field (such as the strip in $D = 2$ or the plate 
in $D = 3$) from those which are {\em irrelevant\/} 
and for which Eq.\,(\ref{A170}) applies.
Thus, the disc in $D = 2$ and the sphere in 
$D = 3$ represent irrelevant perturbations whereas the cylinder 
in $D = 3$ represents a relevant perturbation. This implies that 
an infinitely elongated cylinder in $D = 3$ generates a perturbation
of the order parameter from its bulk value whose spatial extension 
is only limited by the bulk 
correlation length. The order parameter profile
becomes even independent of $R$
in the formal limit $R \to 0$, i.e., if the cylinder 
radius $R$ becomes microscopically small \cite{line}.
The critical adsorption 
transition on such a microscopically thin `needle' 
is characterized by critical exponents which need not be equal to 
the corresponding exponents for the bulk or the half-space 
(see Appendix \ref{app_ex}).
For the disc in $D = 2$ and the sphere in $D = 3$, in contrast, 
the deviation of the order parameter from its bulk value vanishes 
in the limit $R \to 0$. This is reflected by Eq.\,(\ref{A170}) 
which is $-$ apart from the condition in Eq.\,(\ref{A220}) $-$
only valid if the small radius $R$ is still large on the 
microscopic scale. 

%
\unitlength1cm
%
%
The line $d = D - (\beta / \nu)(D)$ itself 
corresponds to marginal perturbations leading to a behavior 
which in general is 
different from Eq.\,(\ref{A170}). We shall neither 
discuss this nor the crossover from marginal behavior to the
behavior described by Eq.\,(\ref{A170}) which may arise for 
points closely above the line. The line $d = D - (\beta / \nu)(D)$
includes, in particular, the generalized cylinder with 
$d = 3$ and $D = 4$ (see Sec.\,\ref{sec_mf}). The marginal 
behavior of this particular generalized cylinder, however, 
is not typical for spheres or cylinders in $D = 3$, which
are strictly irrelevant or relevant perturbations, respectively.
Therefore in the following we shall generally speak of spheres 
if $d = D$ and speak of cylinders if $d = 2$ and $D \ge 3$
(see Fig.\,\ref{diagram}).

We now turn to the consequences of Eq.\,(\ref{A170}) and the 
related properties of
the universal scaling functions ${\cal C}_{+}(\gamma)$ and 
$P_{\pm}(x_{\pm}, y_{\pm})$ defined in Eqs.\,(\ref{A80}) and
(\ref{A10}). First, we consider the sphere, i.e., $d = D$, for
which Eq.\,(\ref{A170}) applies. In order to obtain
the asymptotic behavior of ${\cal C}_{+}(\gamma)$ for
$\gamma = s / R \to \infty$ one basically replaces 
$\langle \Phi({\bf r}) \rangle_{t \, = \, 0}$ 
in the numerator on the 
rhs of Eq.\,(\ref{A80}) by the {\em bulk\/}
two-point correlation function 
$\langle \Phi({\bf r}) \Phi(0) \rangle_{b, \, t \, = \, 0}$
where $r$ can be replaced by $s$ in leading order. 
By using Eqs.\,(\ref{A190}) and (\ref{A200}) in 
conjunction with   
${\cal E}_{D,D} = A_{\,\uparrow}^{\Phi} / B_{\Phi\,}$
one finds
\begin{equation} \label{A230}
{\cal C}_{+}(\gamma \to \infty) \, \to \, c_+ \, 
\Big(\frac{\gamma}{2} \Big)^{- \beta / \nu} \, \, ,
\quad d = D \, \, .
\end{equation}
This checks with the exact result for ${\cal C}_{+}(\gamma)$ 
in Eq.\,(\ref{A85}). For $t \gtrless 0$ 
and $y_{\pm} = R / \xi_{\pm} \to 0$ with
$x_{\pm} = s / \xi_{\pm} > 0$ fixed
the same steps as above lead to
\begin{equation} \label{A240}
P_{\pm}(x_{\pm}, y_{\pm} \to 0) \, - \, P_{\pm, \, b} \, \to \,
c_{\pm} \, (2 y_{\pm})^{\beta / \nu} \, x_{\pm}^{\, - 2 \beta / \nu}
\, {\cal F}_{\pm}(x_{\pm}) \, \, , \quad d = D \, \, .
\end{equation}
Here ${\cal F}_{\pm}$ is the {\em bulk\/} universal 
scaling function defined by (compare Appendix \ref{app_neu})
\begin{equation} \label{A250}
\langle \Phi({\bf r}) \Phi(0) \rangle_{b, \, t}^{\,C} \, = \,
B_{\Phi\,} r^{- \, 2 \beta / \nu} \, {\cal F}_{\pm}(r / \xi_{\pm})
\end{equation}
where the superscript $C$ denotes the cumulant of the correlation 
function.
Equations (\ref{A250}) and (\ref{A200}) imply
the normalization ${\cal F}_{\pm}(0) = 1$.
According to the definition of $\xi_{\pm}$ as the true
correlation length (compare the discussion below Eq.\,(\ref{A20}))
one has 
$\protect{{\cal F}_{\pm}(x_{\pm} \to \infty)} \sim \exp(- x_{\pm})$.
Equation (\ref{A240}) implies that $P_{\pm}(x_{\pm}, y_{\pm} \to 0)$ 
decays to its bulk value with the power law 
$\sim y_{\pm}^{\, \beta / \nu}$ for {\em any\/} value of $x_{\pm}$.

\begin{figure}[t]
\begin{picture}(16,10)
\put(1.5,-4.1){
\setlength{\epsfysize}{14cm}
\epsfbox{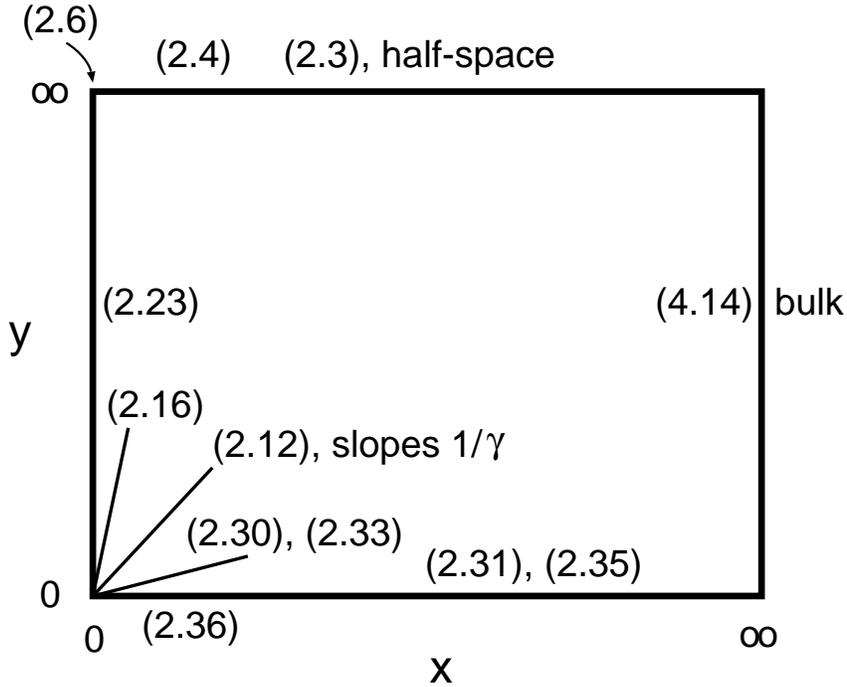}}
\end{picture}
\caption{The universal scaling functions 
$P_{\pm}(x_{\pm} = s / \xi_{\pm}, y_{\pm} = R / \xi_{\pm})$
exhibit distinct behaviors for various limits of the 
variables $x_{\pm}$ and $y_{\pm}$.
The position of an equation number in the $(x,y)$-plane 
indicates the corresponding limiting case to which it applies. 
The case $y_{\pm} = \infty$ corresponds to the 
half-space (see Eq.\,(\ref{A40}) and Table \ref{tab_ca}). 
The short distance behavior for $s \to 0$ corresponds 
to the limit $x_{\pm} \to 0$ with $y_{\pm}$ fixed 
(see Eq.\,(\ref{A150})). 
The universal functions ${\cal C}_{\pm}(\gamma = s / R)$
in Eq.\,(\ref{A65}) characterize the behavior for
$\xi_{\pm} \to \infty$,
i.e., the limit $x_{\pm} \to 0$ with fixed `slope' 
$\gamma^{-1} = y_{\pm} / x_{\pm}$.
The behavior for small $R$ corresponds to the limit
$y_{\pm} \to 0$ with $x_{\pm}$ fixed
(see Eqs.\,(\ref{A240}) and (\ref{A280}))}
\label{fig_land}
\end{figure}

Next, we consider the cylinder, i.e., $d = 2$ and $D \ge 3$.
In accordance with the discussion above this object
represents a relevant perturbation. The order 
parameter deviates from its bulk value even for $R \to 0$ so 
that the universal scaling functions ${\cal C}_{+}(\gamma)$ and 
$P_{\pm}(x_{\pm}, y_{\pm})$ remain finite in the limit 
$\gamma = s / R \to \infty$ and $y_{\pm} = R / \xi_{\pm} \to 0$, 
respectively. For $t = 0$ this implies
\begin{equation} \label{A260}  
{\cal C}_{\pm}(\gamma \to \infty) \, \to \, n_{\pm} \, \, ,
\quad d = 2, \, D \ge 3 \, \, ,
\end{equation}
which defines new universal amplitudes $n_{\pm}$ with 
$n_{-} / n_{+} = ( \xi_{0}^{+} / \xi_{0}^{-} )^{\beta/\nu}$.
Equations (\ref{A260}) 
and (\ref{A62}) imply that for $t = 0$
the order parameter profile near a thin `needle' 
at distances $s$ from the needle large compared with microscopic 
lengths takes the form 
\begin{equation} \label{A270}  
\langle \Phi({\bf r}) \rangle_{t \, = \, 0} \, = \, 
a \, n_{+} \, (s / \xi_{0}^{+})^{- \beta/\nu} \, \, , 
\quad \text{`needle'} \, \, .
\end{equation}
For $t \gtrless 0$ one finds
\begin{equation} \label{A280}
P_{\pm}(x_{\pm}, y_{\pm} \to 0) \, \to \, P_{\pm}(x_{\pm}, 0) 
\, \equiv \, N_{\pm}(x_{\pm}) \, \, ,
\quad d = 2, \, D \ge 3 \, \, ,
\end{equation}
where the new universal scaling functions $N_{\pm}(x_{\pm})$
characterize the critical adsorption profile on a thin 
needle. In the limit $x_{\pm} \to 0$ they behave as
\begin{equation} \label{A290}
N_{\pm}(x_{\pm} \to 0) \, \to \, n_{\pm} 
\, x_{\pm}^{\, - \beta / \nu} \, \, , \quad \text{`needle'} \, \, ,
\end{equation}
with $n_{\pm}$ from Eq.\,(\ref{A260})
(compare Eq.\,(\ref{A40}) for the half-space).
In Sec.\,\ref{sec_mf} we shall confirm 
Eqs.\,(\ref{A260}) - (\ref{A290}) and calculate $n_{\pm}$
and $N_{\pm}(x_{\pm})$ explicitly for $D = 4$.
Figure \ref{fig_land} summarizes the various types of
limiting behavior of the scaling functions 
$P_{\pm}(x_{\pm}, y_{\pm})$.

%
\unitlength1cm
%

\newpage

\section{Excess adsorption}
\label{sec_excess}

\subsection{General scaling properties}
\label{subsec_general}

Close to $T_c$ the total enrichment of the preferred component 
of the fluid near the surface of the generalized 
cylinder is proportional to the excess adsorption $\Gamma(t, R)$
defined as
\begin{eqnarray}
\Gamma(t, R) \, & = & \, \int\limits_{V} d^{D} r \, 
\Big[ \langle \Phi({\bf r}) \rangle_{t} \, - \,
\langle \Phi \rangle_{t, \, b} \Big] \label{E10}\\
& = & \, l^{\delta} \, \Omega_d \, 
\int\limits_{0}^{\infty} ds \, (s + R)^{d - 1} \,
\Big[ \langle \Phi({\bf r}) \rangle_{t} \, - \,
\langle \Phi \rangle_{t, \, b} \Big] \, \, . \nonumber
\end{eqnarray}
Here $V = {\Bbb R}^{D} \setminus K$ is the volume accessible to the
critical fluid, i.e., the total space except for the volume 
occupied by the sphere or the cylinder $K$, and  
$\Omega_{d} = 2 \pi^{d/2}/\,\Gamma(d/2)$ is the surface 
area of the $d$-dimensional unit sphere.
For a sphere one has $\delta = D - d = 0$ 
and $l^{\delta} = 1$ whereas for
a cylinder $d = 2$ so that in $D = 3$ the quantity 
$l^{\delta} = l$ is the length of the cylinder.
In order to obtain the scaling behavior of
$\Gamma(t, R)$ we split the $s$ integration into
the intervals $0 \le s \le \sigma$ and $s > \sigma$ where
$\sigma$ represents a typical microscopic length \cite{floter}. 
For the second interval one can use Eq.\,(\ref{A10}) which 
leads to
\begin{equation} \label{E20}
\Gamma(t, R) \, = \, l^{\delta} \, \Omega_d \, 
\Big[I(t; \sigma, R) \, + \, 
J_{\pm}(y_{\pm}; \sigma / \xi_{\pm}) \Big]
\end{equation}
with 
\begin{equation} \label{E30}
I(t; \sigma, R) \, = \, 
\int\limits_{0}^{\sigma} ds \, (s + R)^{d - 1} \,
\Big[ \langle \Phi({\bf r}) \rangle_{t} \, - \,
\langle \Phi \rangle_{t, \, b} \Big] \, \, ,
\end{equation}
\begin{equation} \label{E40}
J_{\pm}(y_{\pm}; \sigma / \xi_{\pm}) \, = \,
a \, |t|^{\beta} \, \xi_{\pm}^{\,d} \, 
\int\limits_{\sigma / \xi_{\pm}}^{\infty} dx_{\pm} \,
(x_{\pm} + y_{\pm})^{d - 1} \, 
\Big[ P_{\pm}(x_{\pm}, y_{\pm}) - P_{\pm, \, b} \Big] \, \, .
\end{equation}
In the limit $t \to 0$ the integral in Eq.\,(\ref{E30}) remains 
finite and yields a nonuniversal constant which is subdominant
to the diverging contribution $J_{\pm}$ in Eq.\,(\ref{E40}).
In order to clarify the dependence of the latter 
on $y_{\pm} = R / \xi_{\pm}$ we decompose it according to
\begin{eqnarray}
& & J_{\pm}(y_{\pm}; \sigma / \xi_{\pm}) \, = \, 
R^{d-1} \, a \, \xi_{0}^{\pm} \, |t|^{\beta - \nu} \label{E50}\\[1mm]
& & \quad \times \, \Bigg\{
\int\limits_{\sigma / \xi_{\pm}}^{\infty} dx_{\pm} \,
\Big[ P_{\pm}(x_{\pm}) - P_{\pm, \, b} \Big] 
\, + \, G_{\pm}(y_{\pm}; \sigma / \xi_{\pm}) \Bigg\} \nonumber
\end{eqnarray}
where $P_{\pm}(x_{\pm}) = P_{\pm}(x_{\pm}, \infty)$ are the 
scaling functions for the half-space and 
\begin{mathletters} \label{E60}
\begin{eqnarray}
& & G_{\pm}(y_{\pm}; \sigma / \xi_{\pm}) \label{E60a}\\[1mm]
& & = \, \int\limits_{\sigma / \xi_{\pm}}^{\infty} dx_{\pm} \,
\Big\{ \Big( \frac{x_{\pm}}{y_{\pm}} + 1 \Big)^{d - 1} \, 
\Big[ P_{\pm}(x_{\pm}, y_{\pm}) - P_{\pm, \, b} \Big] -   
\Big[ P_{\pm}(x_{\pm}) - P_{\pm, \, b} \Big] 
\Big\} \,\, . \nonumber
\end{eqnarray}
The comparison of Eq.\,(\ref{E50}) with Eq.\,(\ref{E20}) shows
that the first term in curly brackets in Eq.\,(\ref{E50}) renders
a contribution of the interval $s > \sigma$ to the 
excess adsorption per unit area 
as if the surface of the particle would be {\em planar\/} 
times the area $A = l^{\delta} \Omega_d R^{d-1}$
of the actually curved surface of a 
sphere or a cylinder. Thus the functions $G_{\pm}$ in  
Eq.\,(\ref{E50}) reflect the deviation of the 
excess adsorption on a curved surface from that on 
a planar surface beyond pure geometry.

In order to reveal the behavior of 
$J_{\pm}(y_{\pm}; \sigma / \xi_{\pm})$ for $t \to 0$
it is necessary to study the integrals in 
Eqs.\,(\ref{E50}) and (\ref{E60a}) at their lower bounds 
$\sigma / \xi_{\pm} \to 0$. The integral in 
Eq.\,(\ref{E50}) can be analyzed along the 
lines of Sec.\,IIA in Ref.\,\cite{floter}. 
One finds that for $D < 4$ its contribution to 
$\Gamma(t \to 0, R)$ leads to the power law 
singularity $\sim |t|^{\beta - \nu} / (\nu - \beta)$ 
corresponding to a planar surface. 
The proper limit $D \nearrow 4$, for which $\nu \to \beta$
(see Eq.\,(\ref{A45})), is accomplished by the presence of 
a term constant with respect to $t$ which also diverges 
for $D \nearrow 4$ such that the sum 
leads to a contribution
to $\Gamma(t \to 0, R)$ which
diverges logarithmicly in $D = 4$. The integral in 
Eq.\,(\ref{E60a}), however, remains finite for 
$\sigma / \xi_{\pm} \to 0$
also in $D = 4$ since the singular behavior 
for $x_{\pm} \to 0$
of the first term in curly brackets
is cancelled by the second term. This implies that 
the function 
\begin{equation} \label{E50b}
G_{\pm}(y_{\pm}) \, \equiv \, G_{\pm}(y_{\pm}; \, \sigma / \xi_{\pm} = 0)
\end{equation}
does indeed represent the leading behavior for $t \to 0$. 
Note that $G_{\pm}(y_{\pm})$ is universal because it depends only 
on the universal scaling functions $P_{\pm}(x_{\pm}, y_{\pm})$
and $P_{\pm}(x_{\pm})$. In sum one finds
\end{mathletters}
\begin{equation} \label{E70}
\Gamma(t \to 0, R) \, \to \, A \, a \, \xi_{0}^{\pm} \,
\Big\{ g_{\pm} \, \frac{|t|^{\beta - \nu} - 1}{\nu - \beta} 
\, + \, |t|^{\beta - \nu} \, 
G_{\pm}(y_{\pm}) \Big\} \, \, , \quad D \le 4 \, \, ,
\end{equation}
with the universal numbers (see Table\,\ref{tab_excess})
\begin{equation} \label{E80}
g_{\pm} \, = \, \left\{ \begin{array}{l@{\,\,\,}l}
(\nu - \beta) \, {\displaystyle \int\limits_{0}^{\infty}} dx_{\pm}
\,\Big[ P_{\pm}(x_{\pm}) - P_{\pm, \, b} \Big] 
\, \, , \quad & D < 4 \, \, , \\
\, \nu \, c_{\pm} \, \, , \quad & D = 4 \, \, , \end{array} \right.
\end{equation}
where the second line is the limit for $D \nearrow 4$ of the first line.
According to the above discussion one has $G_{\pm}(\infty) = 0$. 
Equation (\ref{E70}) generalizes the corresponding 
Eq.\,(2.10) in Ref.\,\cite{floter} for a planar surface by the
additional second term in Eq.\,(\ref{E70}).
A quantity accessible to experiments is the ratio 
\begin{mathletters} \label{E90}
\begin{equation} \label{E90a}
{\cal R}_{\Phi}(|t|, R) \, = \, 
\frac{\Gamma(+|t|, R)}{\Gamma(-|t|, R)}
\end{equation} 
of the excess adsorptions above and below the 
critical point. The leading behavior of ${\cal R}_{\Phi}$
for $|t| \to 0$ is characterized by a universal
function $R_{\Phi}(|t|, y_{+}, y_{-})$, i.e., 
\begin{equation} \label{E90b} 
{\cal R}_{\Phi}(|t| \to 0, R) \, \to \, 
R_{\Phi}(|t|, y_{+}, y_{-}) \, \, ,
\end{equation}
which can be read off from Eqs.\,(\ref{E90a}) and (\ref{E70}).
For a planar surface this function reduces to the 
universal number \cite{floter}
\end{mathletters}
\begin{equation} \label{E100}
R_{\Phi} \, = \,
\frac{\xi_{0}^{+}}{\xi_{0}^{-}} \,
\frac{g_{+}}{g_{-}} \, \, \, , \quad \text{half-space} \, \, .
\end{equation}
Table\,\ref{tab_excess} summarizes theoretical results 
for $g_{\pm}$ and $R_{\Phi}$ corresponding to the 
half-space. For the curved surface 
of a sphere or cylinder, however, we shall see
that for $y_{\pm} = R / \xi_{\pm} \to 0$
the divergence of the second term in curly 
brackets in Eq.\,(\ref{E70}) is more pronounced than the 
divergence of the first term.
In order to clarify this aspect we now discuss the behavior 
of $G_{\pm}(y_{\pm})$ in the limit for large 
and small values of $y_{\pm}$, respectively.\\[2mm]

%
\begin{table}[tbp]
\medskip
\caption{Numerical values of the universal numbers $g_{\pm}$ 
and $R_{\Phi}$ (see Eqs.\,(\ref{E80}) and (\ref{E100})).}
\label{tab_excess}
\bigskip
\begin{tabular}{lccc} 
$D$ & $g_{+}$ & $g_{-}$ & $R_{\Phi}$ \\ \hline
$4$     & $\sqrt{2}/2 \simeq 0.707$  & $1$ & $1$ \\
$3$, MC\tablenote{Reference \cite{slsl97}} & $0.663$  & $0.599$ \\
$3$, RG$^{\text{a}}$ & $0.581$  & $0.438$ \\ 
$3$, interpolation\tablenote{Reference \cite{floter}; 
obtained from
interpolating between the results in $D = 4 - \varepsilon$ and $D = 2$}
& $0.69 \pm 0.1$ & $0.56 \pm 0.1$ & $2.28 \pm 0.1$ \\ 
$2$     & $0.910$  & $0.0818$ & $22.236$ \\ 
\end{tabular}
\end{table}



\subsection{The scaling function
$G_{\pm}(y_{\pm})$ for $y_{\pm} \to \infty$}
\label{subsec_big}

For $y_{\pm} \gg 1$ we assume that $G_{\pm}(y_{\pm})$  
is analytic in $y_{\pm}^{\,-1} = \xi_{\pm} / R$ 
so that it can be expanded into a Taylor series around 
$y_{\pm}^{\,-1} = 0$, i.e., 
\begin{equation} \label{E110}
G_{\pm}(y_{\pm} \to \infty) \, = \, a_{d,D}^{\,\pm} \, y_{\pm}^{\,-1}
\, + \, b_{d,D}^{\,\pm} \, y_{\pm}^{\,- 2} \, + \, \ldots
\end{equation}
(recall $G_{\pm}(\infty) = 0$ by definition).
The coefficients $a_{d,D}^{\,\pm}$ and $b_{d,D}^{\,\pm}$  
are dimensionless and universal and depend on the space
dimension $D$ and on the
shape of the generalized cylinder,
i.e., on $d$. The validity of the expansion (\ref{E110})
is plausible since in the limit $\xi_{\pm} / R \ll 1$ the thickness 
$\sim \xi_{\pm}$ of the adsorbed layer is much smaller than the
particle radius $R$ so that a small curvature expansion 
should be applicable to $\Gamma(t \to 0, R)$ in Eq.\,(\ref{E70})
in which a surface term 
$\sim l^{\delta} R^{d - 1}$ is followed by successive terms 
$\sim R^{d - 2}$, $R^{d - 3}$, etc.,  
generated by the surface curvature. The terms on the rhs
of Eq.\,(\ref{E110}) correspond to the leading curvature 
contributions to this expansion and imply that
for decreasing values of $y_{\pm} = R / \xi_{\pm} \sim |t|^{\nu}$
the leading corrections to the first term in curly brackets 
in Eq.\,(\ref{E70}) $-$ corresponding to a flat surface $-$
are of the order $|t|^{\beta - \nu\, } |t|^{- n \nu}$ with 
$n = 1, 2$, etc., and start to dominate it as soon 
as $y_{\pm} \lesssim 1$.

Similar as for the short distance expansion of the 
order parameter (compare Eq.\,(\ref{A110})) the first 
Taylor coefficients of the expansion of 
$G_{\pm}(y_{\pm} \to \infty)$ 
also determine the curvature parameters of a particle ${\cal K}$ 
of {\em more general shape\/} provided its surface $S$ is smooth 
and all principal radii of curvature are much larger than 
$\xi_{\pm}$. In this case one expects an expansion of the 
general form
\begin{eqnarray}
& & \Gamma(t \to 0, R) \, = \, a \, \xi_{0}^{\pm}
\label{E120}\\[1mm]
& & \times \, \int\limits_{S} dS \, \Big\{
g_{\pm} \, \frac{|t|^{\beta - \nu} - 1}{\nu - \beta} \, + \,
\lambda_1^{\,\pm} K_m \, + \, \lambda_2^{\,\pm} K_m^{\,2} \, + \,
\lambda_G^{\,\pm} K_G \, + \, \ldots \, \Big\}
\nonumber
\end{eqnarray}
where the curvatures $K_m$ and $K_G$ are given for a 
$(D - 1)$-dimensional surface of general shape
in Eq.\,(\ref{A100}). For the special case that the 
particle ${\cal K}$ is a generalized cylinder $K$
the curvatures $K_m$ and $K_G$ are given by Eq.\,(\ref{A140})
and the comparison of Eq.\,(\ref{E120}) with Eqs.\,(\ref{E70}) 
and (\ref{E110}) yields 
\begin{mathletters} 
\label{E125}
\begin{equation} \label{E125b}
\lambda_1^{\,\pm} \, \frac{d-1}{2} \, = \,
|t|^{\beta-\nu} \, a_{d,D}^{\,\pm} \, \xi_{\pm}
\end{equation}
and the relation
\begin{equation} \label{E125c}
\lambda_2^{\,\pm} \, \frac{(d-1)^2}{4} \, + \,
\lambda_G^{\,\pm} \, \frac{(d-1)(d-2)}{2} \, = \, 
|t|^{\beta-\nu} \, b_{d,D}^{\,\pm} \, \xi_{\pm}^{\,2} \, \, .
\end{equation}
The curvature parameters 
$\lambda_1^{\,\pm}$, $\lambda_2^{\,\pm}$,
and $\lambda_G^{\,\pm}$ depend on $D$ but should not depend on the 
shape of $K$, i.e., on $d$ in Eq.\,(\ref{E125}).
This implies a corresponding dependence on $d$ of 
$a_{d,D}^{\,\pm}$ and $b_{d,D}^{\,\pm}$, which provides
an important consistency check for the validity of
Eq.\,(\ref{E120}).
In Sec.\,\ref{sec_mf} we confirm this dependence and 
explicitly calculate the curvature parameters for $D = 4$. 

\end{mathletters}


\subsection{The scaling function
$G_{\pm}(y_{\pm})$ for $y_{\pm} \to 0$}
\label{subsec_low}

In order to investigate this limit it is convenient to 
consider the function
\begin{eqnarray}
& & y_{\pm}^{\,d - 1} \, G_{\pm}(y_{\pm}) \label{E130}\\[1mm]
& & = \int\limits_{0}^{\infty} dx_{\pm} \,
\Big\{ ( x_{\pm} + y_{\pm})^{d - 1} \, 
\Big[ P_{\pm}(x_{\pm}, y_{\pm}) - P_{\pm, \, b} \Big]
 - y_{\pm}^{\,d - 1} \,
\Big[ P_{\pm}(x_{\pm}) - P_{\pm, \, b} 
\Big] \Big\} \,\, .
\nonumber
\end{eqnarray}
Because for $y_{\pm} \to 0$ the function
$P_{\pm}(x_{\pm}, y_{\pm})$ in Eq.\,(\ref{E130})
behaves qualitatively different for spheres as compared with
cylinders (see Sec.\,\ref{subsec_sre}) we treat 
them separately. 

First, we consider the sphere, i.e., $d = D$, 
for which Eq.\,(\ref{A240}) holds for the behavior of
$P_{\pm}(x_{\pm}, y_{\pm} \to 0)$ if $x_{\pm} > 0$ is
fixed. However, the integral in Eq.\,(\ref{E130}) starts at the 
lower bound $x_{\pm} = 0$ where this condition for $x_{\pm}$ is 
violated. Therefore we temporarily
split the $x_{\pm}$ integration into the 
intervals $0 < x_{\pm} < \sqrt{y_{\pm}}$ and 
$x_{\pm} > \sqrt{y_{\pm}}\,$. In the latter interval 
the relation $s / R > y_{\pm}^{\,-1/2} \to \infty$ holds
so that Eq.\,(\ref{A240}) is applicable.
This leads to (the curly brackets correspond to those in
Eq.\,(\ref{E130}))
\begin{eqnarray}
& & y_{\pm}^{\,d - 1} \, G_{\pm}(y_{\pm}) \, \to \,
\int\limits_{0}^{ \sqrt{y_{\pm}} } 
dx_{\pm} \, \Big\{ \, \, \, \Big\} \label{E140}\\
& & \quad + \, \int\limits_{ \sqrt{y_{\pm}} }^{\infty} dx_{\pm}
\Big[ ( x_{\pm} + y_{\pm})^{D - 1} \, 
\Big[c_{\pm} \, (2 y_{\pm})^{\beta / \nu\,} 
x_{\pm}^{\, - 2 \beta / \nu}
\, {\cal F}_{\pm}(x_{\pm}) \Big] - y_{\pm}^{\,D - 1} \,
\Big[ P_{\pm}(x_{\pm}) - P_{\pm, \, b} \Big]
\Big] \nonumber \, \, .
\end{eqnarray}
In the second integral the variable $y_{\pm}$ in 
$( x_{\pm} + y_{\pm})^{D - 1}$ 
can be replaced by zero in leading order for $y_{\pm} \to 0$
and the term proportional to $y_{\pm}^{\,D - 1}$,
i.e., the half-space contribution,
can be dropped since $D - 1 > \beta / \nu$. 
The resulting integrand is integrable for $x_{\pm} \to 0$
so that the lower bound $\sqrt{y_{\pm}}$ can be replaced 
by zero and one can readily show
that the resulting second integral in Eq.\,(\ref{E140}) 
dominates the first integral for $y_{\pm} \to 0$. 
This leads to the final result
\begin{equation} \label{E150}
G_{\pm}(y_{\pm} \to 0) \, \to \, 
\omega_{\pm\,} \, y_{\pm}^{\, - D + 1 + \beta / \nu} \, \, , 
\quad d = D \, \, ,
\end{equation}
with the universal amplitudes
\begin{equation} \label{E160}
\omega_{\pm} \, = \,
c_{\pm} \, 2^{\beta / \nu} \,
\int\limits_{0}^{\infty} dx_{\pm} \, 
x_{\pm}^{\, D - 1 - 2 \beta / \nu} 
\, \, {\cal F}_{\pm}(x_{\pm}) \, \, .
\end{equation}
The numerical values of $\omega_{\pm}$ for several 
spatial dimensions $D$ as presently available
are summarized in Table \ref{tab_neu}
(see Sec.\,\ref{subsec_exmf} and Appendix \ref{app_neu}).
Equation (\ref{E150}) in conjunction with Eq.\,(\ref{E70}) 
renders the power law divergence
\begin{equation} \label{E165}
\Gamma(t \to 0, R) \, \sim \, |t|^{2 \beta - D \nu}
\, \, , \quad d = D \, \, .
\end{equation}
For $t \to 0$ the universal function $R_{\Phi}(|t|, y_{+}, y_{-})$ 
in Eq.\,(\ref{E90}) approaches the universal number
\begin{equation} \label{E170}
R_{\Phi}(0,0,0) \, = \, 
\lim\limits_{t \to 0} \, \frac{\Gamma(+|t|, R)}{\Gamma(-|t|, R)} 
\, = \, \left(\frac{\xi_{0}^{+}}{\xi_{0}^{-}}\right)^{D - \beta/\nu} \,
\frac{\omega_{+}}{\omega_{-}} \, \, , 
\quad d = D \, \, .
\end{equation}
Whereas the dependence of the function $R_{\Phi}(|t|, y_{+}, y_{-})$
on $y_{+}$, $y_{-}$ is universal but depends on the 
definition used for the correlation length, we note 
that the limit
$R_{\Phi}(0,0,0)$ in Eq.\,(\ref{E170}) is independent of
the definition for the correlation length. The reason
is that according to the first parts of Eqs.\,(\ref{E170}) and 
(\ref{E10}) $R_{\Phi}(0,0,0)$ can be 
expressed in terms of the order parameter profile
$\langle \Phi({\bf r}) \rangle_{t}$ without resorting to
the notion of the correlation length at all.
(The same holds for the number $R_{\Phi}$
in Eq.\,(\ref{E100}) \cite{floter}.)\\[1mm]

%
\begin{table}[tbp]
\medskip
\caption{Numerical values of the universal amplitudes
$\omega_{\pm}$ and $\upsilon_{\pm}$ 
(see Eqs.\,(\ref{E160}) and (\ref{E190}))
as presently available.}
\label{tab_neu}
\bigskip
\begin{tabular}{lcccc} 
$D$ & $\omega_{+}$ & $\omega_{-}$ & $\upsilon_{+}$ & $\upsilon_{-}$ \\ \hline
$4$ & $4 \sqrt{2} \simeq 5.657$  & $8$ & $1.90$ & $1.86$ \\
$3$ & $1.53 \pm 0.05$ & $\simeq 1.47$ & $-$ & $-$ \\
$2$ & $0.515$ & $0.0501$ & not defined &  not defined \\ 
\end{tabular}
\end{table}
%

Next, we consider the cylinder, i.e., $d = 2$ and $D \ge 3$, 
which represents a relevant perturbation 
(see Sec.\,\ref{subsec_sre}). In this case 
Eq.\,(\ref{A280}) holds for the behavior of 
$\protect{P_{\pm}(x_{\pm}, y_{\pm} \to 0)}$. Thus 
$y_{\pm}^{\,d-1\,} G_{\pm}(y_{\pm}) = y_{\pm\,} G_{\pm}(y_{\pm})$
becomes independent
of $y_{\pm}$ in the limit $y_{\pm} \to 0$ and tends to the 
constant given by the integral in Eq.\,(\ref{E130}) with 
$y_{\pm}$ replaced by zero. This leads to
\begin{equation} \label{E180}
G_{\pm}(y_{\pm} \to 0) \, \to \, 
\upsilon_{\pm\,} \, y_{\pm}^{\,- 1} \, \, , 
\quad d = 2 \, , \, \, D \ge 3 \, \, ,
\end{equation}
with the universal numbers
\begin{equation} \label{E190}
\upsilon_{\pm} \, = \, 
\int\limits_{0}^{\infty} dx_{\pm} \, x_{\pm} \, 
\Big[N_{\pm}(x_{\pm}) - P_{\pm, \, b} \Big] \, \, .
\end{equation}
Numerical values of $\upsilon_{\pm}$ are only
known for $D = 4$ at present 
(see Table \ref{tab_neu} and Sec.\,\ref{subsec_exmf}).
Note that the numbers $\upsilon_{\pm}$ are not defined for $D = 2$
(compare Sec.\,\ref{subsec_sre}).
The exponent in the power law in Eq.\,(\ref{E180}) 
differs from that in Eq.\,(\ref{E150}) 
$-$ apart from the difference generated on purely
geometric grounds $-$ by $\beta / \nu$ due
to the different small $y_{\pm}$ behavior of 
the critical adsorption profiles for spheres as compared with 
cylinders. Equation (\ref{E180}) in conjunction with 
Eq.\,(\ref{E70}) leads to
\begin{equation} \label{E200}
\Gamma(t \to 0, R) \, \sim \, |t|^{\beta - 2 \nu} \, \, ,
\quad d = 2 \, \, , D \ge 3 \, \, .
\end{equation}
For $t \to 0$ the universal function $R_{\Phi}(|t|, y_{+}, y_{-})$ 
in Eq.\,(\ref{E90}) tends to the universal number
\begin{equation} \label{E210}
R_{\Phi}(0,0,0) \, = \, 
\lim\limits_{t \to 0} \, \frac{\Gamma(+|t|, R)}{\Gamma(-|t|, R)} 
\, = \, \left(\frac{\xi_{0}^{+}}{\xi_{0}^{-}}\right)^{2} \,
\frac{\upsilon_{+}}{\upsilon_{-}} \, \, , 
\quad d = 2 \, \, , D \ge 3 \, \, ,
\end{equation}
with $\upsilon_{\pm}$ given by Eq.\,(\ref{E190}).
Again, the limit $R_{\Phi}(0,0,0)$ is independent of the 
definition for the correlation length
(compare Eq.\,(\ref{E170})).


\section{Mean-field theory}
\label{sec_mf}

The critical fluctuations of the fluid
are described by the standard Hamiltonian \cite{binder,diehl}
\begin{equation} \label{M10}
{\cal H}\{ \Phi \} \, = \,  
\int\limits_V dV \, \Big\{ \frac{1}{2} (\nabla \Phi)^2
\, + \, \frac{\tau}{2} \, \Phi^2
\, + \, \frac{u}{24} \, \Phi^4 \Big\}
\end{equation}
for a scalar order parameter field $\Phi({\bf r})$
supplemented by the boundary condition $\Phi = + \infty$
at the surface of the sphere or the cylinder corresponding
to the critical adsorption fixed point \cite{ds93}. The 
position vector ${\bf r} \in {\Bbb R}^{D}$ covers the volume 
$V = {\Bbb R}^{D} \setminus K$ accessible to the critical fluid.
The parameter $\tau$ is proportional to $t = (T - T_c) / T_c$
and $u$ is the $\Phi^4$ coupling constant.
The thermal average $\langle \Phi({\bf r}) \rangle$
corresponding to the Hamiltonian in Eq.\,(\ref{M10}) with
the boundary condition for critical adsorption at the 
surface of $K$ can be systematically expanded in terms of 
increasing powers of $u$, i.e.,
\begin{equation} \label{M20}
\langle \Phi({\bf r}) \rangle \, = \,
\sqrt{\frac{6}{u}} \, \Big[m(s; R,\tau) 
\, \, + \, {\cal O}(u) \Big]
\, \, , \quad s = r_{\perp} - R \, \, .
\end{equation}
The leading contribution to this expansion 
corresponds to the mean-field result 
for the order parameter profile which becomes exact in the limit 
$D \nearrow 4$. The profile $m(s; R,\tau)$ is determined by 
minimizing ${\cal H} \{ \Phi \}$ which leads to the 
Ginzburg-Landau type equation of motion
\begin{equation} \label{M40}
m''(s) \, + \, \frac{d-1}{s+R} \, m'(s) \, - \tau \, m(s) \, = \,
m^{3}(s) \, \, ,
\end{equation}
which is supplemented by the boundary conditions \cite{binder,diehl}
\begin{mathletters} \label{M70}
\begin{equation} \label{M70a}
m(s \to 0; R,\tau) \, \to \, \frac{\sqrt{2}}{s} \, \, ,
\end{equation}
\begin{equation} \label{M70b}
m(s \to \infty; R,\tau) \, \to \, m_{\pm, \, b} \, = \,
\left\{ \begin{array}{l@{\quad}l}
{\displaystyle{0 \, \, \, , } \atop } & 
{\displaystyle{\tau \ge 0 \, \, ,} \atop } \\[1mm]
|\tau|^{1/2} \, \, \, , & \tau < 0 \, \, . \end{array} \right. \\[2mm]
\end{equation}
The parameter $\tau$ is related to $t = (T - T_c) / T_c$ and 
to the correlation length $\xi_{\pm}$ by
\smallskip

\end{mathletters}
\begin{equation} \label{M50}
\xi_{\pm} \, = \, \xi_{0}^{\pm} \, |t|^{- 1/2} \, = \,
\left\{ \begin{array}{l@{\quad}l}
{\displaystyle{\tau^{- 1/2} \, \, \, ,} \atop } & 
{\displaystyle{\tau > 0 \, \, ,} \atop } \\
\displaystyle{\frac{1}{\sqrt{2}}} \, \, |\tau|^{- 1/2} \, \, \, , 
& \tau < 0 \, \, , \end{array} \right.
\end{equation}
and (compare the text following Eq.\,(\ref{A20})) 
\begin{equation} \label{M60}
a |t|^{1/2} = \sqrt{\frac{6}{u}} \, \, |\tau|^{1/2} 
\, \, , \quad \tau < 0 \, \, .
\end{equation}

Equations (\ref{M40}) and (\ref{M70}) uniquely determine the 
profile $m(s;R,\tau)$, which is related to the universal scaling 
functions $P_{\pm}(x_{\pm}, y_{\pm})$ in Eq.\,(\ref{A10}) by
\begin{equation} \label{M80}
P_{\pm}(x_{\pm}, y_{\pm}) \, = |\tau|^{-1/2} \, m(s; R,\tau) 
\, \, , \quad D = 4 \, \, .
\end{equation}
Within the present mean-field approach this scaling 
form holds because the lhs of Eq.\,(\ref{M80}) 
is dimensionless so that it depends only on the dimensionless 
variables $s |\tau|^{1/2}$ and $R |\tau|^{1/2}$, where $|\tau|^{1/2}$ 
is related to $\xi_{\pm}$ by Eq.\,(\ref{M50}). Similarly, 
the universal scaling function ${\cal C}_{+}(\gamma)$ 
in Eq.\,(\ref{A62}) is given by
\begin{equation} \label{M90}
{\cal C}_{+}(\gamma) \, = \, s \, m(s; R,\tau = 0)
\, \, , \quad D = 4 \, \, .
\end{equation}
Equations (\ref{M80}) and (\ref{M90}) lead to exact 
results for $D \nearrow 4$ and generalized cylinders $K$
with arbitrary $d$ in the interval $1 \le d \le D$ 
(compare Fig.\,\ref{diagram}).
Equations (\ref{M40}) and (\ref{M70}) can be solved numerically, 
e.g., by means of a shooting method \cite{shoot}. In some cases 
analytical solutions are available.
In the remaining part of this section we present the
corresponding explicit results for the problems discussed in
Secs.\,\ref{sec_op} and \ref{sec_excess}.


\subsection{Order parameter profiles}
\label{subsec_opmf}

\begin{figure}[t]
\caption{(a) Scaling function $P_{+}(x_+, y_+)$ for a sphere in 
mean-field approximation (i.e., $d = D$ with $D = 4$) 
as function of $x_{+}$ for several values of $y_{+}$. 
The curves start at the value $c_{+}$ (compare the inset in
Fig.\,\ref{fig_hs}). The curve for $y_{+} = \infty$ corresponds 
to the half-space profile (see Eq.\,(\ref{A30})). The curves 
decrease with
decreasing $y_{+}$ and vanish in the limit $y_{+} \to 0$. 
The slopes at $x_+ = 0$ can be read off from 
Eq.\,(\ref{M150}) with $s / R = x_+ / y_+$. 
(b) Same representation for $P_{-}(x_-, y_-)$. The function 
$U(x_{-}, y_{-})$ introduced for convenience
is defined in Eq.\,(\ref{M100}).}
\begin{picture}(16,20)
\put(-0.5,12.8){
\setlength{\epsfysize}{11cm}
\epsfbox{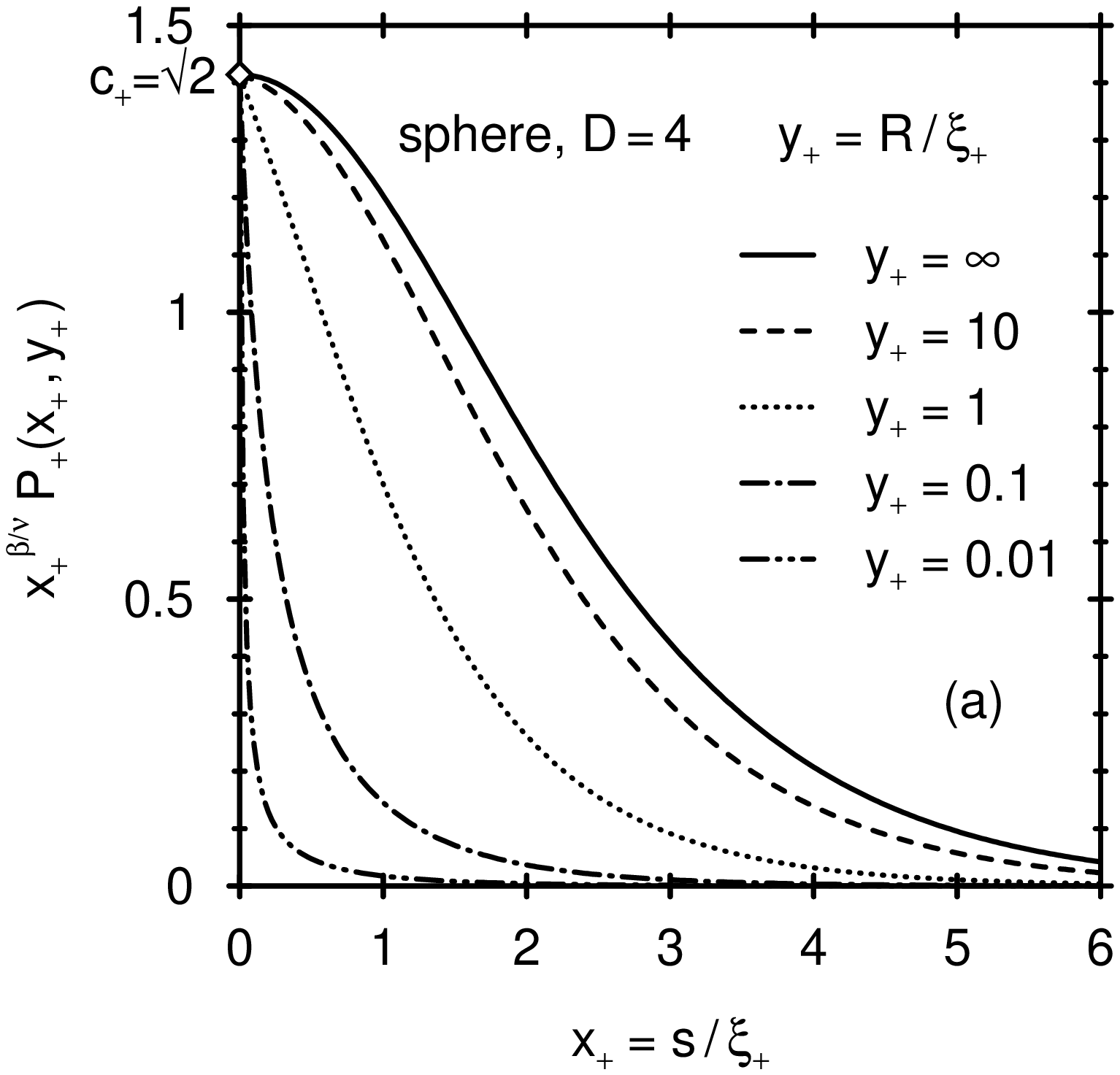}}
\put(-0.5,0.8){
\setlength{\epsfysize}{11cm}
\epsfbox{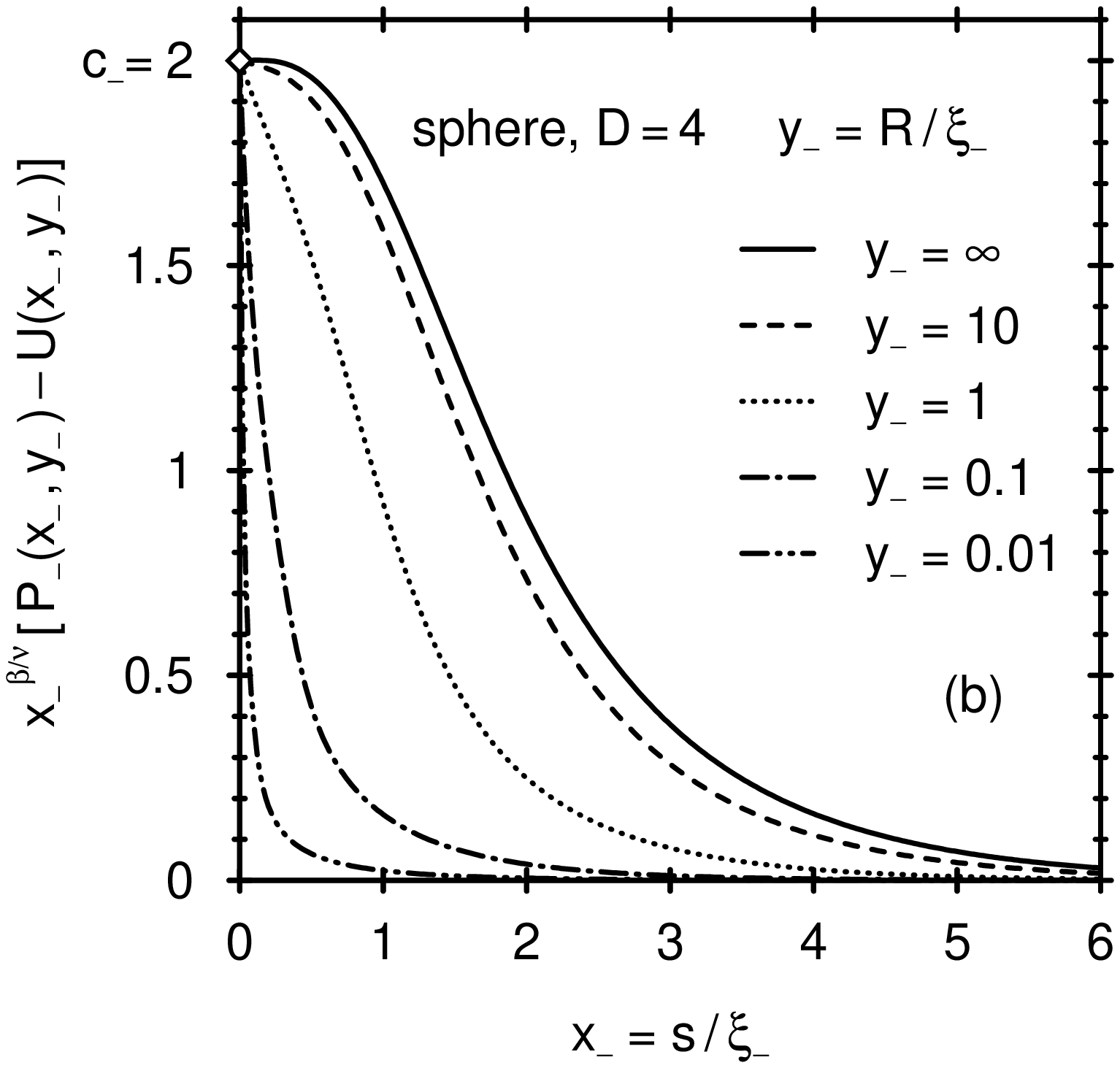}}
\end{picture}
\label{kugel}
\end{figure}

We start with the universal scaling functions $P_{\pm}(x_{\pm},y_{\pm})$
according to Eqs.\,(\ref{M80}) and (\ref{M40}). Figure\,\ref{kugel}
shows their behavior as function of $x_{\pm} = s / \xi_{\pm}$ 
for various values of the parameter $y_{\pm} = R / \xi_{\pm}$
in the case of a sphere (i.e., $d = D$ with $D = 4$, compare 
Sec.\,\ref{subsec_sre} and Fig.\,\ref{diagram}). 
For our presentation we choose
a scaled form which reflects the behavior of 
$P_{\pm}(x_{\pm} \to 0,y_{\pm})$ (compare the inset of 
Fig.\,\ref{fig_hs}; here $\beta / \nu = 1$).
Accordingly, the curves start at the mean-field values 
$c_{+} = \sqrt{2}$ and $c_{-} = 2$ (see Eq.\,(\ref{A150}) 
and Table\,\ref{tab_ca}) , respectively, and decay 
exponentially for $x_{\pm} \to \infty$.\\[2cm]

%
\unitlength1cm
%

\newpage

%
\unitlength1cm
\begin{figure}[t]
\begin{picture}(16,20)
\put(-0.5,12.8){
\setlength{\epsfysize}{11cm}
\epsfbox{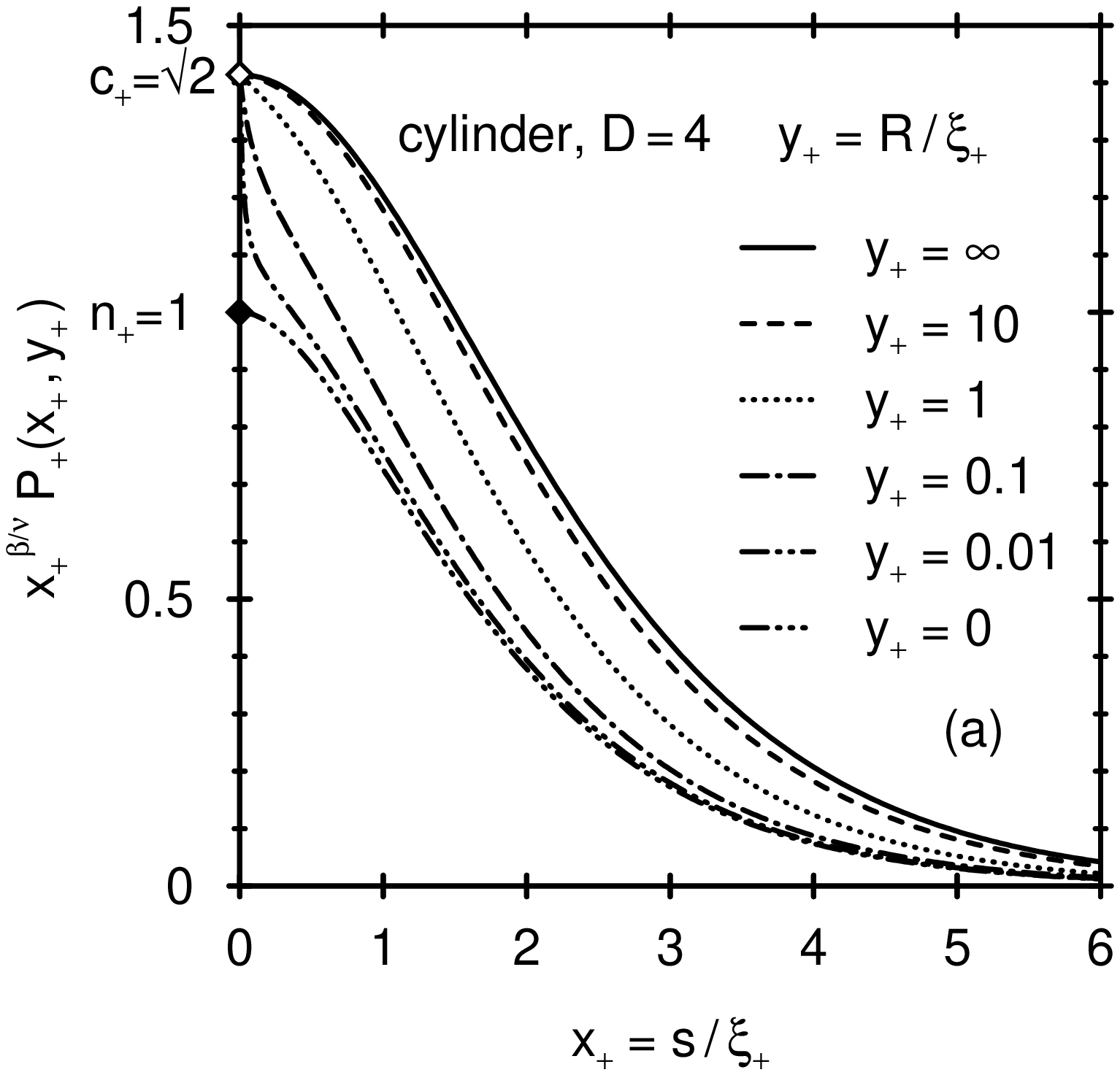}}
\put(-0.5,0.8){
\setlength{\epsfysize}{11cm}
\epsfbox{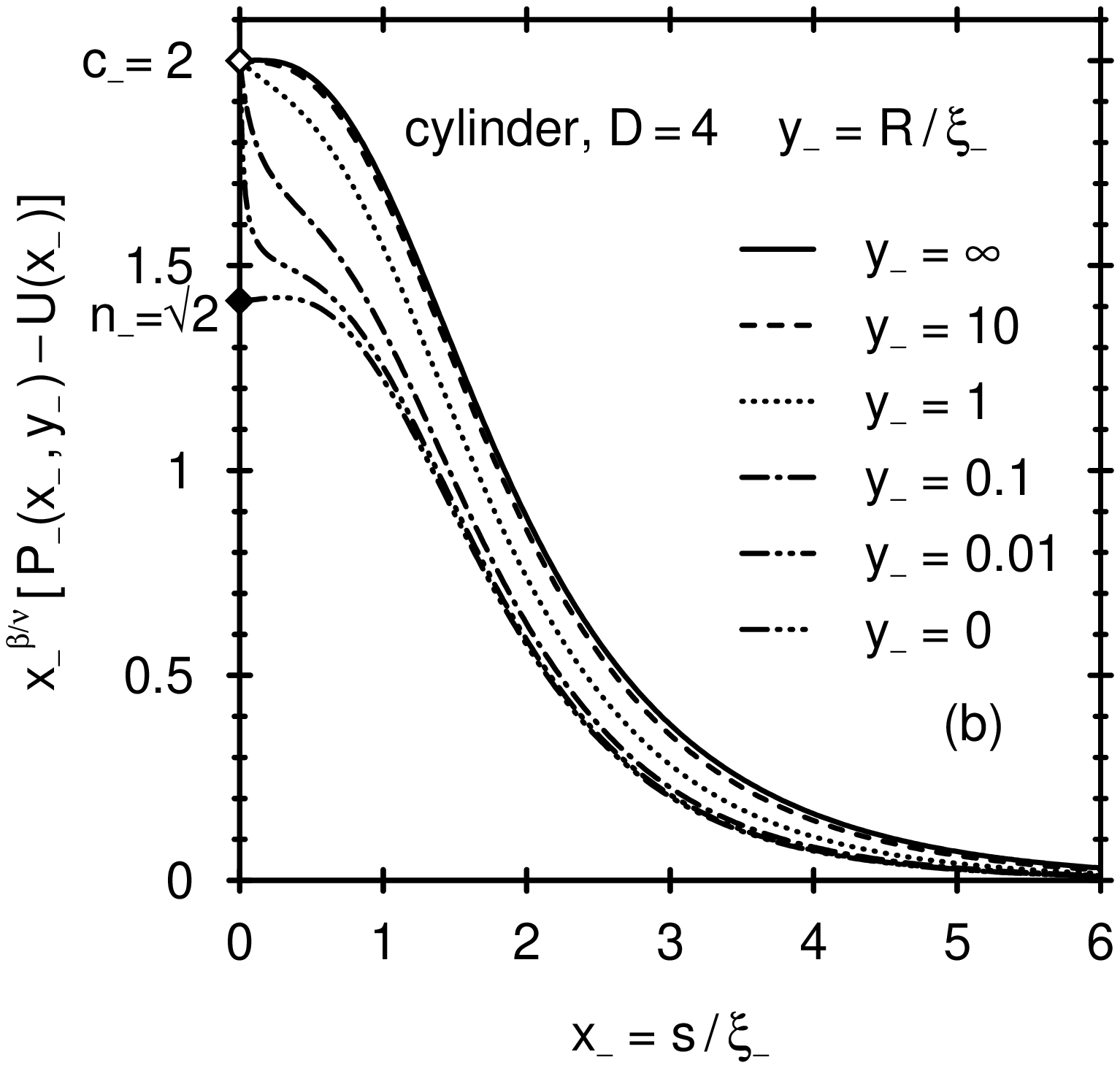}}
\end{picture}
\caption{(a) Scaling function $P_{+}(x_+, y_+)$ for a cylinder in 
mean-field approximation (i.e., $d = 2$ and $D = 4$) 
as function of $x_{+}$ for several values 
of $y_{+}$ (compare Fig.\,\ref{kugel}). The curves decrease 
with decreasing $y_{+}$. In the limit $y_{+} \to 0$
they tend to the curve corresponding to 
$P_{+}(x_{+},0) = N_{+}(x_{+})$ describing the 
critical adsorption profile
on a thin needle. The latter curve starts at the value $n_{+}$ 
(filled symbol). (b) Same representation for $P_{-}(x_-, y_-)$. The 
function $U(x_{-})$ equals $U(x_{-}, \infty)$ from Eq.\,(\ref{M100}).}
\label{cyl}
\end{figure}
%

\medskip

\noindent
In order to deal with 
the rapid decay of $P_{-}(x_{-} \to \infty, y_{-}) - 1$
for small values of $y_{-}$ in the case $T < T_c$
we replace the function $U(x_{-})$ in Eq.\,(\ref{A61}) by
\begin{equation} \label{M100}
U(x_{-}, y_{-}) \, = \, \tanh\Big(
\frac{x_{-}^{\,2}}{x_{-} + 1} \, \, \frac{y_{-} + 1}{y_{-}} \Big)
\, \, , \quad d = D = 4 \, \, ,
\end{equation}
so that $U(x_{-}, \infty) = U(x_{-})$. We emphasize again
that the function $U(x_{-}, y_{-})$ is introduced only 
in order to facilitate an
appropriate representation of $P_{-}(x_{-}, y_{-})$
which reflects the behavior for both small and large values of $x_{-}$ 
for all values of $y_{-}$ (compare the discussion related
to Fig.\,\ref{fig_hs}). The overall dependence
of $P_{\pm}(x_{\pm}, y_{\pm})$ on the parameter $y_{\pm}$
is in line with Eq.\,(\ref{A240}) 
and the related discussion. Starting from the 
half-space behavior for $y_{\pm} = \infty$ the profiles
$P_{\pm}(x_{\pm}, y_{\pm}) - P_{\pm, \, b}$
decrease with decreasing $y_{\pm}$ for any 
fixed value of $x_{\pm}$
and vanish in the limit $y_{\pm} \to 0$.

This latter behavior for a sphere differs from the corresponding 
one for a cylinder (i.e., $d = 2$ and $D = 4$, 
compare Sec.\,\ref{subsec_sre} and Fig.\,\ref{diagram}).
Figure\,\,\ref{cyl} shows the corresponding behavior of
$P_{\pm}(x_{\pm}, y_{\pm})$ for various values of $y_{\pm}$.
For the presentation of $P_{-}(x_{-}, y_{-})$ in the case 
of the cylinder the function 
$U(x_{-}) = U(x_{-}, \infty)$ is suitable 
for all values of $y_{-}$.
The overall dependence on $y_{\pm}$ is characterized by the 
fact that $P_{\pm}(x_{\pm}, y_{\pm})$ does not vanish in the
limit $y_{\pm} \to 0$ but rather tends to the finite limit 
function $P_{\pm}(x_{\pm}, 0) = N_{\pm}(x_{\pm})$
corresponding to the critical adsorption profile on
a thin needle. This is in line with Eq.\,(\ref{A280}) 
and the related discussion. According to Eq.\,(\ref{A290})
the curves for $y_{\pm} = 0$ start at the mean-field values 
of the universal amplitudes $n_{\pm}$ which are given by
\begin{equation} \label{M110}
n_{+} = 1 \, \, , \quad n_{-} = \sqrt{2} \, \, , \qquad D = 4 \, \, .
\end{equation}

\begin{figure}[t]
\begin{picture}(16,12)
\put(-0.5,4){
\setlength{\epsfysize}{11cm}
\epsfbox{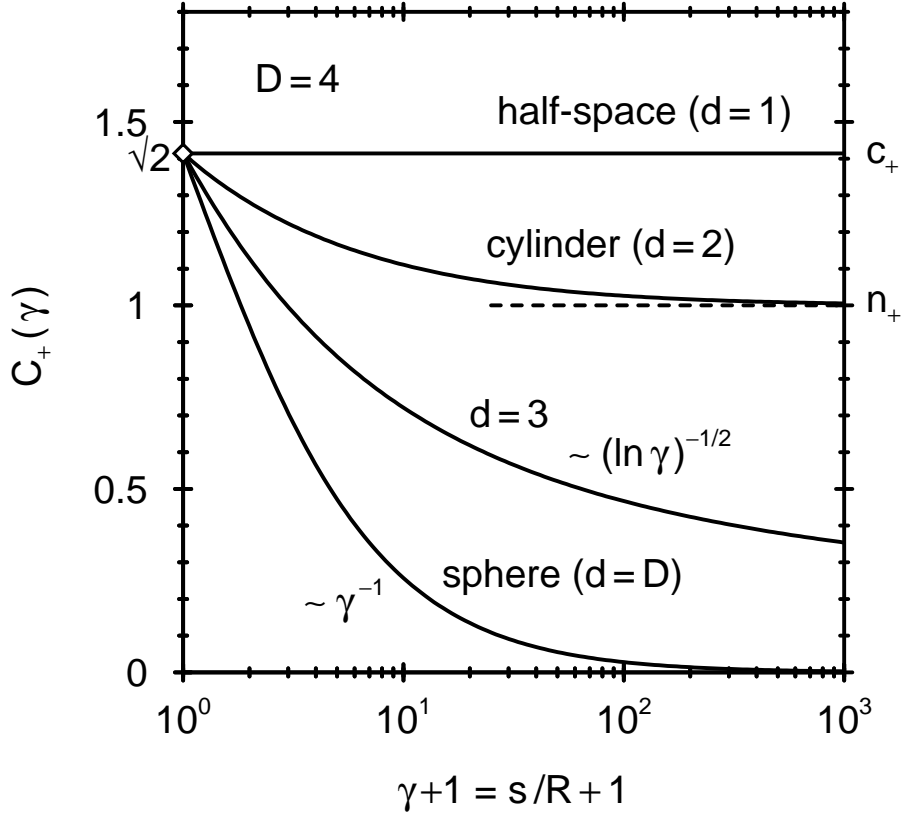}}
\end{picture}
\caption{Scaling function ${\cal C}_{+}(\gamma)$ in mean-field 
approximation (i.e., $D = 4$) as function of $\gamma = s/R$ for several 
values of $d$ (compare Fig.\,\ref{diagram}). All curves
start at the same value ${\cal C}_{+}(0) = c_{+}$ corresponding 
to the half-space but for large $\gamma$ the behaviors are
qualitatively different. For a sphere 
${\cal C}_{+}(\gamma \to \infty)$ vanishes as $\gamma^{-\beta/\nu}$
with $\beta/\nu = 1$
whereas for a cylinder it tends to the finite value $n_{+}$. 
In the marginal case $d = 3$ it vanishes only logarithmically.}
\label{amp}
\end{figure}

Next, we consider the universal scaling function
${\cal C}_{+}(\gamma)$ in Eq.\,(\ref{M90}) corresponding
to $\tau = 0$. Figure\,\ref{amp} shows its behavior for
the present case $D = 4$ and several values of
$d$. The function ${\cal C}_{+}(\gamma)$ starts at
${\cal C}_{+}(0) = c_{+}$ (see the text following 
Eq.\,(\ref{A65})) and for $\gamma \to \infty$ it vanishes 
in the case of the sphere (i.e., $d = D$ with $D = 4$) 
while it tends to the finite
%
%
\unitlength1cm
%
%
\noindent
number $n_{+}$ from Eq.\,(\ref{M110}) in the case of 
the cylinder (i.e., $d=2$). In the marginal case $d = 3$ it 
vanishes only logarithmically (compare 
Sec.\,\ref{subsec_sre} and Fig.\,\ref{diagram}).
For $\tau = 0$ analytical solutions for Eq.\,(\ref{M40}) are 
available in some special cases even for $d > 1$.
For example, for $\tau = 0$ and $d = 4$ 
the differential equation can be solved
by relating it to the generalized Emden-Fowler
equation \cite{gnutzmann}, or by realizing that it 
represents a so-called `equidimensional' equation
(see, e.g., section 1.4 in Ref.\,\cite{bender}).
The result is 
\begin{equation} \label{M112}
m(s; R,\tau = 0) \, = \, \frac{2 \sqrt{2} \, R}{s \, (s + 2 R)}
\, \, \, , \qquad d = 4 \, \, , 
\end{equation}
which agrees with Eqs.\,(\ref{M90}) and (\ref{A85}). 
It is worthwhile to note that an analytical solution
also exists for the particular case $\tau = 0$ and 
$d = 5/2$ for which
\begin{equation} \label{M114}
m(s; R,\tau = 0) \, = \, \frac{\sqrt{2} \, / \, 2}
{s + R - \sqrt{R\,(s+R)}}
\, \, \, , \qquad d = 5/2 \, \, ,
\end{equation}
from which the corresponding result for ${\cal C}_{+}(\gamma)$
can be inferred from Eq.\,(\ref{M90}). Finally, for $\tau = 0$,
$R = 0$, and arbitrary values of $d < 3$ one finds 
\begin{equation} \label{M116}
m(s; R = 0, \tau = 0) \, = \, \frac{\sqrt{3 - d}}{s} 
\, \, \, , \qquad s = r_{\perp} \, \, , \, \, d < 3 \, \, ,
\end{equation}
so that ${\cal C}_{+}(\infty) = n_+ = \sqrt{3 - d} > 0$ for $d < 3$.
This is in line with Eq.\,(\ref{M114}) as well as with 
Fig.\,\ref{diagram} and the related discussion,
according to which in $D = 4$ a generalized cylinder $K$ with 
$d < 3$ represents a relevant perturbation. In particular, 
Eq.\,(\ref{M116}) leads to $m(r_{\perp}) = 1 / r_{\perp}$ 
representing the mean-field profile near a thin needle 
(for which $d = 2$, compare Sec.\,\ref{subsec_sre}) 
at criticality. In Appendix\,\,\ref{app_ex} we calculate the 
corresponding two-point correlation function in the presence of 
this profile. 

The asymptotic behavior of $P_{\pm}(x_{\pm}, y_{\pm})$
for $D = 4$ in various limiting 
cases can be derived directly from the defining 
Eqs.\,(\ref{M80}) and (\ref{M40}) 
even for those cases in which no full analytical solution 
is available \cite{upton}. 

(i) $x_{\pm} = s / \xi_{\pm} \gg 1$ with 
$y_{\pm} = R / \xi_{\pm}$ fixed. 
Because in this limit
$m(s;R,\tau) - m_{\pm, \, b}$ is
exponentially small, in Eq.\,(\ref{M40}) one can
neglect powers of it larger than one.
This leads to a linear differential equation for 
$m(s;R,\tau) - m_{\pm, \, b}$ which can be solved
in terms of modified Bessel functions 
$K_{\alpha}(x_{\pm} + y_{\pm})$ and 
$I_{\alpha}(x_{\pm} + y_{\pm})$ \cite{abr}
with index $\alpha = (d - 2) / 2$. Here only the decreasing
function $K_{\alpha}$ must be considered. Using the asymptotic
behavior of $K_{\alpha}$ for large arguments \cite{abr}
and Eq.\,(\ref{M80}) one finds
\begin{equation} \label{M120} 
P_{\pm}(x_{\pm}, y_{\pm}) - P_{\pm, \, b} \, \to \, 
A_{\pm}(y_{\pm}; d) \, (x_{\pm} + y_{\pm})^{- \frac{d-1}{2}} \,
\exp(-x_{\pm}) \, \, , \quad x_{\pm} \gg 1 \, \, ,
\end{equation}
where the amplitude-function $A_{\pm}(y_{\pm}; d)$ remains 
undetermined by the present method and must be evaluated 
numerically in general. An important feature of Eq.\,(\ref{M120})
is that for increasing values of $x_{\pm}$ the exponential 
decay is enhanced by the algebraic prefactor. Thus the decay 
is weaker if $x_{\pm} \ll y_{\pm}$ and it is weaker for a 
cylinder ($d = 2$) than for a sphere ($d = D$ with $D = 4$).
This is expected because the
perturbation of the bulk fluid is strongest in case of the
half-space (corresponding to $x_{\pm} \ll y_{\pm}$),
less for a cylinder, and even less for a sphere.

(ii) $x_{\pm} \to 0$ with $R$ and $\xi_{\pm}$ fixed, 
i.e., $s \to 0$ (compare Sec.\,\ref{subsec_short}).
The leading behavior of $m(s \to 0;R,\tau)$ can be obtained by
inserting the ansatz
\begin{equation} \label{M130} 
m(s;R,\tau) \, = \, \frac{\sqrt{2}}{s} \, + \, a_0 \, + \,
a_{1\,} s \, + \, {\cal O}(s^2)
\end{equation}
into Eq.\,(\ref{M40}) and fixing the coefficients $a_0$ and $a_1$
such that the prefactors of the most singular powers of $s$ cancel. 
The resulting expression can be arranged so that it takes the 
form implied by Eqs.\,(\ref{A150}) and (\ref{M80}), i.e.,
\begin{eqnarray}
& & m(s \to 0;R,\tau) \, =  \, \frac{\sqrt{2}}{s} \,
\Big\{ 1 \, - \, \frac{1}{3} \, \frac{d-1}{2} \, \frac{s}{R}
\label{M150}\\[2mm]
& & \, + \, \Big[ \frac{5}{9} \, \frac{(d-1)^2}{4} \, - \,
\frac{1}{3} \, \frac{(d-1)(d-2)}{2} \Big] \frac{s^2}{R^2} \,
\, - \, \frac{1}{6} \tau s^2 \, + \, \ldots \Big\} \, \, .
\nonumber
\end{eqnarray}
Equation\,(\ref{M150}) confirms that the leading dependence 
on $\tau$ is analytic with the same prefactor as for the 
half-space. In addition, one can read off the 
curvature parameters according to mean-field theory, i.e.,
\begin{equation} \label{M160}
\kappa_1 = - \, \frac{1}{3} \, \, , \quad
\kappa_2 = \frac{5}{9} \, \, , \quad 
\kappa_G = - \, \frac{1}{3} \, \, , \qquad D = 4 \, \, .
\end{equation}
We note that the derivation of Eq.\,(\ref{M150}) implicitly
contains a consistency check for the validity of the general
curvature expansion (\ref{A150}). For example, the parameters
$\kappa_2$ and $\kappa_G$ are fixed by considering only
two different types of generalized cylinders $K$ with
nonvanishing curvature, e.g., those for which $d = 2$ and 
$d = 3$. Equation (\ref{M160}), however, holds for any value 
of $d$ in the interval $1 \le d \le D$ with $D = 4$ which 
encompasses, in particular, the three integer values 
$d = 2$, $3$, and $4$ (compare Fig.\,\ref{diagram}). 
Thus the curvature parameters are overdetermined. 
In the present case, however, this consistency may be
regarded as a simple consequence of the differential 
equation (\ref{M40}) in which the perturbation generated
by the surface curvature is proportional to $d - 1$.
In the next subsection we consider 
the small curvature expansion (\ref{E120}) for the excess 
adsorption, for which the corresponding consistency check
in $D = 4$ provides a more stringent test.

(iii) $y_{\pm} \to 0$ with $s$ and $\xi_{\pm}$ fixed, i.e., $R \to 0$
(compare Sec.\,\ref{subsec_sre}). We consider the case of a 
sphere, i.e., $d = D$ with $D = 4$. By inserting 
the ansatz $m(s; R, \tau) - m_{\pm,\,b} = R \, u(s; \tau)$ into 
Eq.\,(\ref{M40}) and keeping only terms linear in $R$ one 
obtains a linear differential equation for $u(s; \tau)$, 
similar as in case (i) above. The solution of this
differential equation in terms 
of the modified Bessel function $K_1(x_{\pm})$ \cite{abr} in 
conjunction with Eq.\,(\ref{M80}) leads to 
\begin{equation} \label{M170}
P_{\pm}(x_{\pm}, y_{\pm} \to 0) \, - \, P_{\pm, \, b} \, \to \,
2 c_{\pm} \, y_{\pm} \, x_{\pm}^{\,-1} \, K_1(x_{\pm}) 
\, \, , \quad d = D = 4 \, \, .
\end{equation}
Here the constant prefactor $2 c_{\pm}$ is fixed because the limit 
$x_{\pm} \to 0$ of the rhs of Eq.\,(\ref{M170}) must 
reproduce Eq.\,(\ref{A65}), in which
${\cal C}_{\pm}(\gamma \to \infty) \to 2 c_{\pm} / \gamma$ for
$d = D = 4$ according to Eq.\,(\ref{A85}).
Equation (\ref{M170}) is in line with Eq.\,(\ref{A240}) 
because for $D = 4$ one has $\beta/\nu = 1$ and
the universal scaling functions
${\cal F}_{\pm}(x_{\pm})$ are given by
\begin{equation} \label{M180}
{\cal F}_{\pm}(x_{\pm}) \, = \, x_{\pm\,} K_1(x_{\pm}) 
\, \, , \qquad D = 4 \, \, .
\end{equation}


\subsection{Excess adsorption}
\label{subsec_exmf}

For $D = 4$ Eq.\,(\ref{E70}) reduces to
\begin{equation} \label{Q10}
\Gamma(t \to 0, R) \, \to \, A \, a \, \xi_{0}^{\pm} \,
\Big\{ g_{\pm\,} |\ln|t|\,| \, + \,
G_{\pm}(y_{\pm}) \Big\} \, \, , \quad D = 4 \, \, ,
\end{equation}
with $g_{+} = 1 / \sqrt{2}$ and $g_{-} = 1$. The universal
scaling function $G_{\pm}(y_{\pm})$ in Eq.\,(\ref{Q10}) 
is given by Eq.\,(\ref{E60}) with
$P_{\pm}(x_{\pm}, y_{\pm})$ and $P_{\pm}(x_{\pm})$
from Eqs.\,(\ref{M80}) and (\ref{A30}).
In the following we discuss the behavior of $G_{\pm}(y_{\pm})$
in the limit of large and small values of 
$y_{\pm} = R / \xi_{\pm}$
and give results for the whole range of $y_{\pm}$
representing the crossover between these two limits.  

(i) $G_{\pm}(y_{\pm})$ for $y_{\pm} \to \infty$
(compare Sec.\,\ref{subsec_big}). By using 
the differential equation (\ref{M40}) in conjunction
with Eq.\,(\ref{M80}) one can determine the coefficients
$a_{d,D}^{\,\pm}$ and $b_{d,D}^{\,\pm}$ in Eq.\,(\ref{E110})
for generalized cylinders $K$
with $D = 4$ and arbitrary $d$ in the interval $1 \le d \le 4$. 
This calculation is presented in Appendix \ref{appB}.
As a result one finds that the dependences of 
$a_{d,4}^{\,\pm}$ and $b_{d,4}^{\,\pm}$ on $d$ are
precisely of the form given by Eq.\,(\ref{E125}). 
In the present case $D = 4$ one has 
$|t|^{\beta - \nu} = 1$ and 
the curvature parameters are
given by \cite{interior}
\begin{mathletters} \label{Q20}
\begin{equation} \label{Q20a}
\lambda_1^{\,+} \, = \, 5.09 \, \, \xi_+ \, \, , \, \, \, 
\lambda_1^{\,-} \, = \, 4.91 \, \, \xi_- \, \, ,
\end{equation}
\begin{equation} \label{Q20b}
\lambda_2^{\,+} \, = \, - \, 1.55 \, \, \xi_+^{\,2} \, \, , \, \, \,
\lambda_2^{\,-} \, = \, - \, 0.91 \, \, \xi_-^{\,2} \, \, ,
\end{equation}
\begin{equation} \label{Q20c}
\lambda_G^{\,+} \, = \, 2.87 \, \, \xi_+^{\,2} \, \, , \, \, \, 
\lambda_G^{\,-} \, = \, 2.52 \, \, \xi_-^{\,2} \, \, .
\end{equation}
The consistency with Eq.\,(\ref{E125}) 
for $D = 4$ can be traced back to
nontrivial properties of the differential equation 
(\ref{M40}) (see Appendix \ref{appB}) and thus provides a 
valuable and important check for the validity of the 
general curvature expansion (\ref{E120}). We expect 
that the curvature expansion (\ref{E120}) is also valid in 
$D = 3$ and that the corresponding
numerical prefactors multiplying $\xi$ or $\xi^2$ in the 
curvature parameters for $D = 3$ differ only quantitatively 
from those in Eq.\,(\ref{Q20}) valid for $D = 4$.   

\end{mathletters}

As an illustration, consider a curved {\em membrane\/} 
with both sides exposed to a fluid near criticality. 
In particular we consider the {\em total\/} excess 
adsorption, i.e., the sum of the excess adsorptions 
on each side of the membrane, per unit area. 
In this case the contributions to $\lambda_1^{\,\pm} K_m$ 
in the expansion (\ref{E120}) from each side cancel and 
the signs in Eqs.\,(\ref{Q20b}) and (\ref{Q20c}) in conjunction 
with Eq.\,(\ref{A140}) imply that the total excess 
adsorption is {\em larger\/} near spherical regions 
($d = D$) of the membrane as compared to flat regions, 
whereas near cylindrical regions ($d = 2$) it is 
{\em smaller\/} as compared to flat regions. 

(ii) $G_{\pm}(y_{\pm})$ for $y_{\pm} \to 0$ (compare 
Sec.\,\ref{subsec_low}). In this case the
behaviors for spheres and cylinders are qualitatively 
different due to the different behaviors of 
$P_{\pm}(x_{\pm}, y_{\pm})$. 
For a sphere (i.e., $d = D$ with $D = 4$)
the power law (\ref{E150}) is valid where the
exponent $- D + 1 + \beta / \nu$ equals $- 2$ and
according to Eqs.\,(\ref{E160}) and (\ref{M180})
the universal amplitudes $\omega_{\pm}$ are given by
\begin{equation} \label{Q30}
\omega_{+} \, = \, 4 \sqrt{2} \, \, , \quad 
\omega_{-} \, = \, 8 \, \, , \qquad d = D = 4 \, \, .
\end{equation}
For a cylinder (i.e., $d = 2$ and $D = 4$) one finds 
the behavior (\ref{E180}) where the universal numbers 
$\upsilon_{\pm}$ in Eq.\,(\ref{E190})
can be evaluated numerically with the result
\begin{equation} \label{Q40}
\upsilon_{+} \, = \, 1.90 \, \, , \quad 
\upsilon_{-} \, = \, 1.86 \, \, , \qquad d = 2, \, D = 4 \, \, .
\end{equation}

\begin{figure}[t]
\caption{Scaling functions $G_{\pm}(y_{\pm})$ for a sphere
in mean-field approximation (i.e., $d = D$ with $D = 4$) 
as function of $y_{\pm}^{\,-1} = \xi_{\pm} / R$ for 
(a) $T > T_c$ and (b) $T < T_c$.
The dotted lines show the small curvature expansion
(see Eqs.\,(\ref{E110}) - (\ref{E125}), and (\ref{Q20}))
valid for $y_{\pm}^{\,-1} \ll 1$
and the dashed lines show the power law 
(see Eqs.\,(\ref{E150}) and (\ref{Q30})) valid for 
$y_{\pm}^{\,-1} \to \infty$; for $D=4$ the exponent 
$-D+2 + \beta/\nu$ is equal to $ - 1$.}
\begin{picture}(16,20)
\put(-0.5,12.8){
\setlength{\epsfysize}{11cm}
\epsfbox{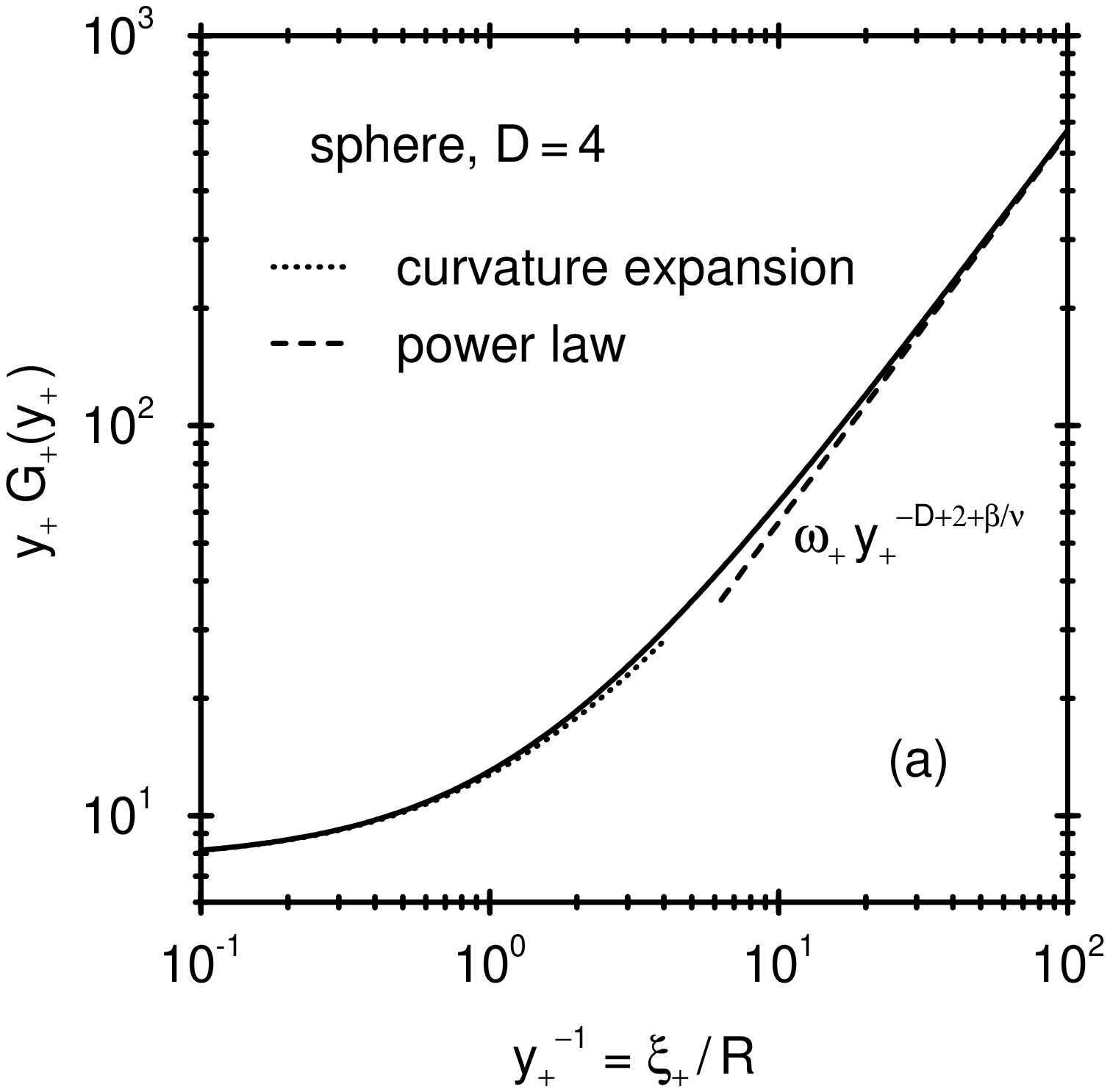}}
\put(-0.5,0.8){
\setlength{\epsfysize}{11cm}
\epsfbox{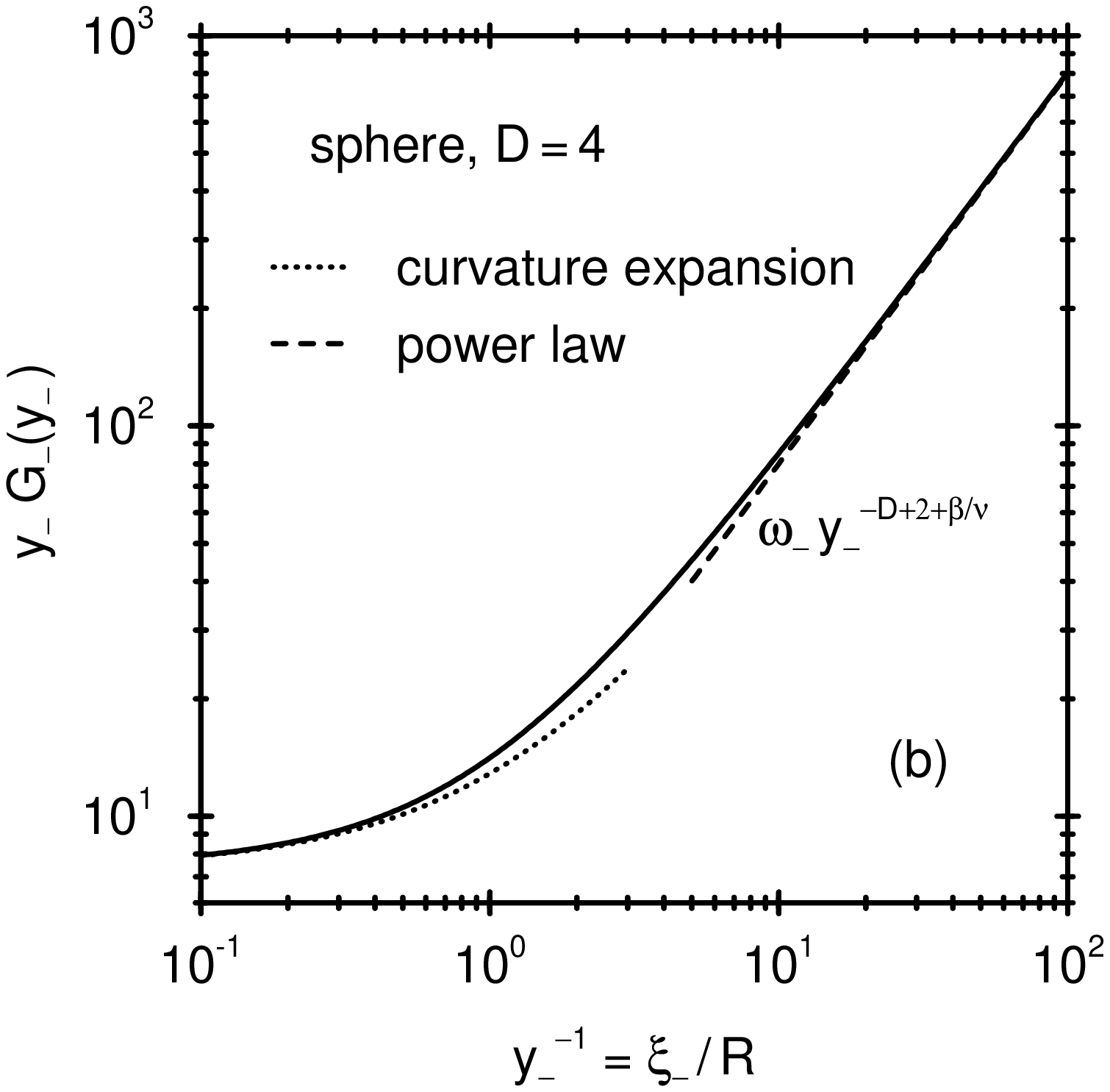}}
\end{picture}
\label{fig7}
\end{figure}

\begin{figure}[t]
\begin{picture}(16,17.5)
\put(-0.5,7.5){
\setlength{\epsfysize}{11cm}
\epsfbox{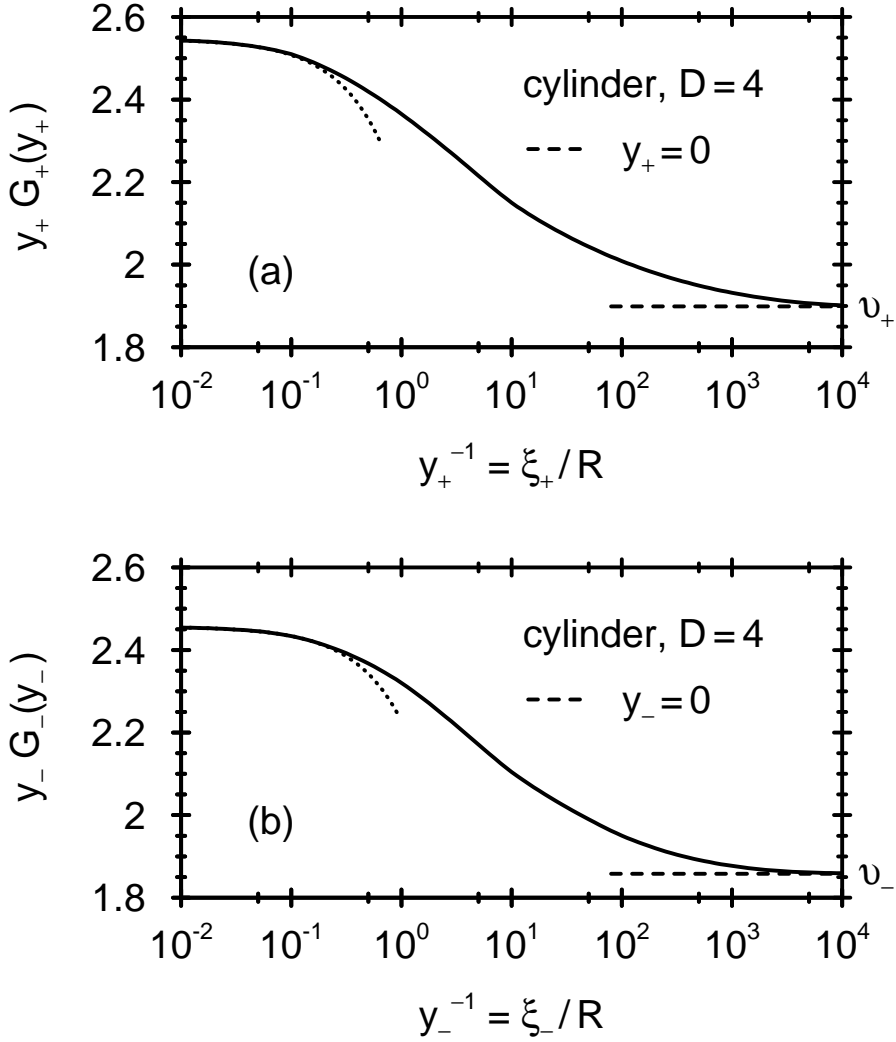}}
\end{picture}
\caption{Scaling functions $G_{\pm}(y_{\pm})$ 
for a cylinder 
in mean-field approximation (i.e., $d = 2$ and $D = 4$) 
as function of $y_{\pm}^{\,-1} = \xi_{\pm} / R$ for 
(a) $T > T_c$ and (b) $T < T_c$ (compare Fig.\,\ref{fig7}).
The dotted lines show the small curvature expansion
and the dashed lines 
correspond to the numbers $\upsilon_{\pm}$ in the behavior 
$G_{\pm}(y_{\pm} \to 0) \to \upsilon_{\pm\,} y_{\pm}^{\,-1}$
(see Eqs.\,(\ref{E180}) and (\ref{Q40})).}
\label{fig8}
\end{figure}

(iii) The full scaling functions $G_{\pm}(y_{\pm})$ 
describe the crossover between their analytic behaviors 
for $y_{\pm} = R / \xi_{\pm} \to \infty$ and 
the power laws for $y_{\pm} \to 0$ 
which have been discussed in (i) and (ii) above.
Figures \ref{fig7} and \ref{fig8} show numerical 
results for $G_{\pm}(y_{\pm})$ corresponding
to a sphere and a cylinder, respectively, in $D = 4$.
The results confirm the small curvature expansion as implied 
by Eqs.\,(\ref{E110}) - (\ref{E125}), and (\ref{Q20}) and 
the properties (\ref{E150}) and (\ref{E180}),
and they provide the range of validity of 
the asymptotic behavior. Note that
in the case of the sphere
$y_{\pm} G_{\pm}(y_{\pm})$ as function of 
$y_{\pm}^{\,-1} = \xi_{\pm} / R$ diverges for 
$y_{\pm}^{\,-1} \to \infty$
with the corresponding power law whereas in the case of the
cylinder with increasing $y_{\pm}^{\,-1}$
it interpolates between 
one finite value related to $\lambda_1^{\,\pm}$ to the 
other finite value $\upsilon_{\pm}$.\\[1mm]
 
%
\unitlength1cm
%

\newpage

%
\unitlength1cm
%

\newpage

\section{Summary and concluding remarks}
\label{summary}

We have studied critical adsorption phenomena on spherical 
and cylindrical particles of radius $R$ which are immersed 
in a fluid near criticality, $t = (T - T_c) / T_c \to 0$,
for the case that the fluid is at 
the critical composition. The corresponding adsorption 
profiles at the radial distance $s$ from the surface
are characterized by universal scaling functions 
$P_{\pm}( s / \xi_{\pm}, R / \xi_{\pm} )$
for $T \not= T_c$, involving 
the bulk correlation lengths $\xi_{\pm}$ for 
$T \gtrless T_c$, and ${\cal C}_{+}(s/R)$ for $T = T_c$ 
(see Eqs.\,(\ref{A10}) and (\ref{A62}), respectively).

\medskip

In the following we summarize our main results starting with 
{\em local properties\/} of 
$P_{\pm}(x_{\pm}, y_{\pm})$ and ${\cal C}_{+}(\gamma)$
in various limiting cases as indicated by Fig.\,\ref{fig_land}.
\begin{enumerate}
\item[(1)]
For $T = T_c$ we have introduced the short
distance expansion of the order parameter profile 
near a weakly curved $(D - 1)$-dimensional surface of 
{\em general\/} shape (see Eq.\,(\ref{A110})). This expansion
involves the local curvature invariants $K_m$, $K_m^{\,2}$,
and $K_G$ (see Eq.\,(\ref{A100})). The corresponding
expansion parameters $\kappa_1$, $\kappa_2$, and $\kappa_G$
appear also in the short distance
expansion (\ref{A150}) of $P_{\pm}(x_{\pm}, y_{\pm})$ 
valid for $T \gtrless T_c$. The parameters 
$\kappa_1$, $\kappa_2$, and $\kappa_G$ are universal and
depend only on the space dimension $D$
(see Eqs.\,(\ref{A120}), (\ref{A130}), and (\ref{M160})). 
\item[(2)]
For $R \ll s, \, \xi$ the order parameter profile near a 
cylinder becomes independent of $R$ in the limit $R \to 0$, i.e., 
if the cylinder radius is microscopically small 
(see Eqs.\,(\ref{A260}) - (\ref{A290})). 
In contrast, near a sphere the `universal part' of the order parameter 
profile, i.e., the one described by Eqs.\,(\ref{A230}) and 
(\ref{A240}), vanishes for $R \to 0$. 
This {\em qualitative\/} difference in behavior 
can be explained by means of a 
small radius operator expansion (see Eqs.\,(\ref{A170}) 
and (\ref{A180})). As indicated by Fig.\,\ref{diagram}
a sphere is an irrelevant perturbation
for the fluid near criticality whereas a
cylinder is a relevant perturbation. 
\item[(3)]
The explicit forms of the universal scaling functions
$P_{\pm}(x_{\pm}, y_{\pm})$ and ${\cal C}_{+}(\gamma)$ 
for a sphere and a cylinder within mean-field approximation
(see Figs.\,\ref{kugel} - \ref{amp}) corroborate
the results described in the above paragraphs (1) and (2).
\end{enumerate}

\medskip

We now turn to the {\em excess adsorption\/} $\Gamma(t \to 0, R)$
describing the total enrichment of the preferred component
of the fluid near criticality in proximity of a sphere or 
cylinder (see Eq.\,(\ref{E10})).
The curvature dependence of $\Gamma(t \to 0, R)$
is characterized by universal scaling functions 
$G_{\pm}(R / \xi_{\pm})$ obtained from   
$P_{\pm}(s / \xi_{\pm}, R / \xi_{\pm} )$ by 
integrating over the first variable $s / \xi_{\pm}$
(see Eqs.\,(\ref{E60}) and (\ref{E70})).
\begin{enumerate}
\item[(4)]
For $R / \xi_{\pm} \gg 1$ we have introduced the
expansion of $\Gamma(t \to 0, R)$ near a weakly curved 
$(D - 1)$-dimensional surface of {\em general\/} shape 
in terms of the local curvature invariants $K_m$, $K_m^{\,2}$,
and $K_G$ (see Eq.\,(\ref{E120})). The corresponding
expansion parameters $\lambda_1^{\,\pm}$, $\lambda_2^{\,\pm}$,
and $\lambda_G^{\,\pm}$ depend only on the space dimension $D$
and can be expressed in terms of the universal coefficients
$a_{d,D}^{\,\pm}$ and $b_{d,D}^{\,\pm}$ appearing in the 
expansion of $G_{\pm}(y_{\pm} \to \infty)$ 
(see Eqs.\,(\ref{E110}) - (\ref{E125})).
The explicit calculation of 
$\lambda_1^{\,\pm}$, $\lambda_2^{\,\pm}$, and $\lambda_G^{\,\pm}$
within mean-field approximation (see Eq.\,(\ref{Q20}) and 
Appendix \ref{appB}) provides an important check for the 
validity of the general curvature expansion of the
excess adsorption.
\item[(5)]
For $R / \xi_{\pm} \to 0$ the behaviors of 
$\Gamma(t \to 0, R)$ and $G_{\pm}(y_{\pm})$
are characterized by power laws  
(see Eqs.\,(\ref{E150}) - (\ref{E165}) for a sphere and  
(\ref{E180}) - (\ref{E200}) for a cylinder). 
The exponents in these power laws are different for spheres 
and cylinders beyond the purely geometrical effects.
This can be traced back
to the corresponding difference in behavior of 
$P_{\pm}(x_{\pm}, y_{\pm})$ addressed in paragraph (2).
\item[(6)]
The explicit forms of $G_{\pm}(y_{\pm})$ for a sphere and
a cylinder in mean-field approximation 
(see Figs.\,\ref{fig7} and \ref{fig8}) confirm the general
results presented in paragraphs (4) and (5).
\end{enumerate}

\medskip

We conclude by summarizing some of the {\em fieldtheoretic
developments\/} and {\em perspectives\/} for 
future theoretical work, simulations, and experiments.
\begin{enumerate}
\item[(7)]
For a systematic field-theoretical analysis of 
the critical adsorption 
phenomena on spheres and cylinders it is useful to introduce 
the particle shape of a `generalized cylinder' 
(see Eq.\,(\ref{I10}) and Fig.\,\ref{fig_cyl}) which is
characterized by the space dimension $D$
and an internal dimension $d$ encompassing a sphere, 
a cylinder, and a planar wall as special cases. 
\item[(8)]
For a single sphere, 
certain asymptotic behaviors of the order parameter {\em profile\/} 
are characterized by universal quantities 
(see Eqs.\,(\ref{A150}) and (\ref{A240})). These 
quantities are accurately known in $D = 3$ 
(see Eq.\,(\ref{A120}), (\ref{A130}), and Table \ref{tab_ca};
as far as the scaling functions ${\cal F}_{\pm}(x_{\pm})$ are
concerned, only the function ${\cal F}_{+}(x_{+})$ appropriate 
for $T > T_c$ is known accurately for $D = 3$, 
see Appendix \ref{app_neu}). 
The corresponding predictions can be tested using small angle
scattering of light, X-rays or neutrons which can probe the
enlarged effective size of the colloidal particles due to
the adsorption.
\item[(9)]
For a single sphere, in $D = 3$ the behavior 
of the {\em excess adsorption\/} in the limit $R / \xi_{\pm} \to 0$
is available as well
(see Eqs.\,(\ref{E70}), (\ref{E150}), and Table \ref{tab_neu})
and can be tested experimentally by, e.g., volumetric measurements.
For a disc in $D = 2$ the numerical values of 
the universal amplitudes $\omega_{\pm}$ 
in Eq.\,(\ref{E150}) are also known (compare Appendix \ref{app_neu}).
This can be relevant for protein inclusions in fluid membranes
\cite{gil97}.
\item[(10)]
Apart from the curvature dependence of the excess 
adsorption (see Eq.\,(\ref{E120})) also the corresponding 
energetic contributions, i.e., the change of the surface 
tension generated by the surface curvature, can be relevant for 
applications. For example, in case of a membrane immersed
in a fluid near criticality such contributions and
their temperature dependence are expected to influence the intrinsic 
bending rigidities of the membrane. These modifications of the
bending rigidities can, in turn, induce shape changes of 
vesicles formed by closed membranes in a controllable way 
(see, e.g., Ref.\,\cite{dobereiner}). 
\item[(11)]
A one-dimensional extended perturbation in an Ising-like system
which breaks the symmetry of the order parameter represents a relevant 
perturbation (see Sec.\,\ref{subsec_sre}) of the bulk system. 
This gives rise to new universal quantities such as 
the critical exponents $\eta_{\parallel}$ and $\eta_{\perp}$
characterizing the decay of the structure factor
(see Appendix \ref{app_ex} 
and in particular Eqs.\,(\ref{G90}) and (\ref{G110}))
and the amplitudes $n_{\pm}$ 
and $\upsilon_{\pm}$ (see Eqs.\,(\ref{A290}) and (\ref{E190}),
respectively; the mean-field values of $n_{\pm}$ 
and $\upsilon_{\pm}$ are quoted in 
Eqs.\,(\ref{M110}) and (\ref{Q40})). 
Apart from a rodlike particle immersed in a fluid, such 
a one-dimensional perturbation could also be realized in a {\em solid\/}, 
e.g., by a lattice dislocation in a binary alloy
if the dislocation locally breaks the symmetry of the order parameter.
This would be a natural starting point for testing the numerous 
fieldtheoretic predictions derived here by Monte Carlo simulation.
\item[(12)]
As stated in the Introduction,
the results addressed in paragraph (2) 
can be relevant for the
flocculation of colloidal particles dissolved in a fluid near 
criticality. Specificly, if one considers the case $T = T_c$ 
and compares the behaviors of the 
order parameter profiles near
a sphere and near an infinitely elongated cylinder,
respectively, 
the order parameter of the fluid 
at a distance $s \gg R$ from the particle with radius $R$
is larger by a factor $\sim (s/R)^{\beta/\nu} \gg 1$ for the 
cylinder than for the sphere
(see Eqs.\,(\ref{A62}), (\ref{A230}) and (\ref{A260})).
For rods with a large but finite
length $l$ and distances $s \gtrsim l$ this enhancement
is of the order of $(l / R)^{\beta/\nu} \gg 1$. 
Since the critical Casimir forces between particles are 
partially induced by an overlap of the order parameter 
profiles generated by the individual particles 
in the space between them, the aforementioned results
suggest that not only spheres but, {\em a fortiori\/},
also long rods in a fluid near criticality 
can aggregate due to 
critical Casimir forces.
\end{enumerate}

\section*{Acknowledgements}

We thank Prof. E. Eisenriegler for helpful discussions
and Profs. A. J. Bray and M. E. Fisher for helpful 
correspondence.
This work has been supported by the 
German Science Foundation through
Sonderforschungsbereich 237 
{\em Unordnung und gro{\ss}e Fluktuationen\/}.


\appendix

\section{Two-point correlation function near a needle} 
\label{app_ex}

The two-point correlation function
$G({\bf r}, {\bf r}\,')$ for Gaussian fluctuations
around the profile $m(r_{\perp}) = 1 / r_{\perp}$ 
(see the text following Eq.\,(\ref{M116})) {\em at\/} 
criticality satisfies the differential equation  
\begin{equation} \label{G10}
\Big[ - \Delta_D \, + \, 3 \, m^{2}(r_{\perp}) \Big] \, 
G({\bf r}, {\bf r}\,') \, = \, 
\delta^{(D)}({\bf r} - {\bf r}\,')
\end{equation}
where $\Delta_D$ is the Laplacian operator in $D$-dimensional
space. (Although the present mean-field calculation implies
$D = 4$ we leave the symbol $D$ for clarity.)
In coordinates adapted to the geometry of a generalized
cylinder $K$ with $d = 2$ and $R = 0$ the Laplacian operator 
$\Delta_D$ has the form
\begin{equation} \label{G20}
\Delta_D \, = \, \Delta_{\perp} \, + \, \sum\limits_{i=1}^{D-2} \, 
\frac{\partial^2}{\partial r_{\parallel, \, i}^{\, \, \, 2}}
\end{equation}
with
\begin{equation} \label{G30}
\Delta_{\perp} \, = \, \frac{\partial^2}{\partial r_{\perp}^{\,2} } \, + \, 
\frac{1}{r_{\perp}} \,
\frac{\partial}{\partial r_{\perp}} \, + \, \frac{1}{ r_{\perp}^{\,2} } \,
\frac{\partial^2}{\partial \vartheta^2}
\end{equation}
as the Laplacian operator 
in the two-dimensional subspace perpendicular to $K$.
Here $\vartheta$ denotes the angle between ${\bf r}_{\perp}$ and
a fixed direction in the radial subspace. In order to 

%
\unitlength1cm
\begin{figure}[t]
\begin{picture}(16,13)
\put(2,-3){
\setlength{\epsfysize}{16cm}
\epsfbox{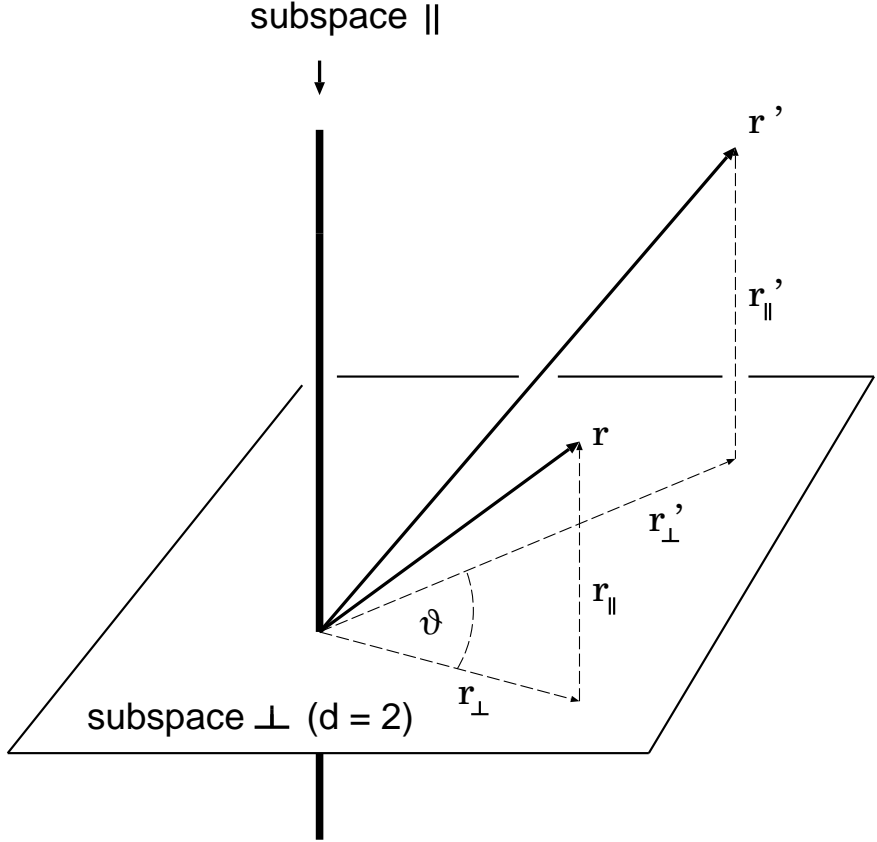}}
\end{picture}
\caption{A thin `needle' corresponding to a generalized cylinder $K$ 
with $d = 2$ and $R = 0$ (compare Eq.\,(\ref{I10}) and 
Fig.\,(\ref{fig_cyl})). The two spatial arguments
${\bf r}$ and ${\bf r}\,'$ of the correlation function
in Eq.\,(\ref{G40}) are also shown.}
\label{fig_needle}
\end{figure}
%
%
\noindent
solve Eq.\,(\ref{G10}) we carry out the Fourier transform 
in the subspace parallel to $K$ and a partial wave decomposition 
in the radial subspace, i.e., 
\begin{eqnarray}
& & G({\bf r}, {\bf r}\,') \, = \,
G(r_{\perp}, r_{\perp}^{\,\,'}\, , \vartheta, 
| {\bf r}_{\parallel} - {\bf r}_{\parallel}^{\,\,'\,}|) 
\label{G40}\\[2mm] 
& & = \, \sum_{n=0}^{\infty} \, W_{n}(\vartheta) \,
\int\frac{d^{D-2\,}p}{(2 \pi)^{D-2}} \,
\exp[i \, {\bf p} \,
({\bf r}_{\parallel} - {\bf r}_{\parallel}^{\,\,'} \,)] \, \,
\widetilde{G}_{n}(r_{\perp},r_{\perp}^{\,\,'}, p) \, \, , \nonumber
\end{eqnarray}
where $\vartheta$ is now the angle between ${\bf r}_{\perp}$ and 
${\bf r}_{\perp}^{\,\,'}$ (see Fig.\,\ref{fig_needle}).
The functions $W_{n}(\vartheta)$ are given by
\begin{equation} \label{G50}
W_{n}(\vartheta) \, = \, 
\frac{2 - \delta_{n, 0}}{2 \pi} \, \, 
\cos (n \vartheta)
\end{equation}
with $\delta_{n, 0} = 1$ for $n = 0$ and zero otherwise. They obey
$\sum_n W_{n}(\vartheta) = \delta(\vartheta) / 2$ where the 
support of the $\delta$-function on the rhs is located entirely 
in the interval $\vartheta \ge 0$ so that 
$\int_0^{\vartheta'} d\vartheta \, \delta(\vartheta) = 1$ 
for any $\vartheta' > 0$.
The propagator $\widetilde{G}_{n}$ in Eq.\,(\ref{G40})
satisfies the radial equation
\begin{equation} \label{G60}
\Big[- \frac{\partial^2}{\partial r_{\perp}^{\,2} } \, - \, 
\frac{1}{r_{\perp}} \,
\frac{\partial}{\partial r_{\perp}} \, + \, p^2 \, + \,
\frac{n^2 + 3}{r_{\perp}^{\,2}} \, \Big] \, 
\widetilde{G}_{n}(r_{\perp},r_{\perp}^{\,\,'}, p) \, = \,
\frac{\delta(r_{\perp} - r_{\perp}^{\,\,'})}{r_{\perp}} \, \, .
\end{equation}
For $r_{\perp} \neq  r_{\perp}^{\,\,'}$ the solutions of 
this equation are 
modified Bessel functions $K_{\lambda}(p \, r_{\perp})$ and 
$I_{\lambda}(p \, r_{\perp})$ \cite{abr} with index
\begin{equation} \label{G70}
\lambda \, = \lambda_n \, = \, \sqrt{n^2 + 3} \, \, .
\end{equation}
The physically acceptable 
solution $\widetilde{G}_{n}$ of Eq.\,(\ref{G60}) is 
uniquely determined by the
requirements that (i) it is symmetric in $r_{\perp}$ and 
$r_{\perp}^{\,\,'}$, (ii) it decays for $r_{\perp} \to \infty$, and 
(iii) it does not diverge as the singularity at $r_{\perp} = 0$ 
in Eq.\,(\ref{G60}) is approached. The result is
\begin{equation} \label{G80}
\widetilde{G}_{n}(r_{\perp},r_{\perp}^{\,\,'}, p) \, = \,
K_{\lambda}(p \, r_{\perp}^{(>)}) \, I_{\lambda}(p \, r_{\perp}^{(<)})
\end{equation}
with $r_{\perp}^{(>)} = \max(r_{\perp}, r_{\perp}^{\,\,'}\,)$,
$r_{\perp}^{(<)} = \min(r_{\perp}, r_{\perp}^{\,\,'}\,)$, and
$\lambda$ given by Eq.\,(\ref{G70}). Equations (\ref{G40}) and 
(\ref{G80}) represent the two-point correlation 
function near a thin needle in the presence of the adsorption profile 
$m(r_{\perp}) = 1  / r_{\perp}$ at criticality. In the following 
we infer the asymptotic behavior of $G({\bf r}, {\bf r}\,')$ in 
various limits. 

(a) $r_{\perp}$, $r_{\perp}^{\,\,'}$ fixed and 
$| {\bf r}_{\parallel} - {\bf r}_{\parallel}^{\,\,'\,}| \to \infty$.
The rhs of Eq.\,(\ref{G80}) can be expanded for 
$\protect{p \to 0}$, e.g., 
by relating $K_{\lambda}$ to $I_{- \lambda}$ and $I_{\lambda}$ via 
the corresponding 
formula in section 9.6.2 of Ref.\,[45(a)] and by using the
formula in section 9.6.10 of the same reference. Apart from terms 
proportional to integer powers of $p^2$, which are analytic in 
${\bf p}$ yielding only exponentially decaying contributions
on the lhs of Eq.\,(\ref{G40}),
one finds that the {\em leading singular\/} contribution 
behaves as $p^{2 \lambda}$. Regarding the leading behavior 
for $| {\bf r}_{\parallel} - {\bf r}_{\parallel}^{\,\,'\,}| \to \infty$
in Eq.\,(\ref{G40})
only the term with $n = 0$ is important so that 
\begin{mathletters}
\label{G90}
\begin{equation} \label{G90a}
G(r_{\perp}, r_{\perp}^{\,\,'}\, , \vartheta, 
| {\bf r}_{\parallel} - {\bf r}_{\parallel}^{\,\,'\,}| \to \infty)
\, \sim \, \frac{1}
{| {\bf r}_{\parallel} - {\bf r}_{\parallel}^{\,\,'\,}|
^{D - 2 + \eta_{\parallel}}}
\end{equation}
with 
\begin{equation} \label{G90b}
\eta_{\parallel} = 2 \lambda_0 = 2 \sqrt{3} \, \, .
\end{equation}

\end{mathletters}

(b) ${\bf r}_{\parallel} = {\bf r}_{\parallel}^{\,\,'\,}$,
$r_{\perp}^{\,\,'}$ fixed and $r_{\perp} \to \infty$. 
For $r_{\perp} > r_{\perp}^{\,\,'}$ using 
Eq.\,(\ref{G40}) and rescaling $x = p \, r_{\perp}$ leads to
\begin{equation} \label{G100}
G(r_{\perp}, r_{\perp}^{\,\,'}\, , \vartheta, 0) \, = \,
\frac{1}{r_{\perp}^{\,D-2}} \, \sum_{n=0}^{\infty} \, 
W_{n}(\vartheta) \, \int\frac{d^{D-2\,}x}{(2 \pi)^{D-2}} \,
K_{\lambda}(x) \, I_{\lambda}(x r_{\perp}^{\,\,'} / r_{\perp}) \, \, .
\end{equation}
Here $I_{\lambda}(x r_{\perp}^{\,\,'} / r_{\perp})$ can be
expanded for $r_{\perp}^{\,\,'} / r_{\perp} \to 0$ so that
in this limit again only the term with $n = 0$ is important 
and 
\begin{mathletters}
\label{G110}
\begin{equation} \label{G110a}
G(r_{\perp} \to \infty, r_{\perp}^{\,\,'}\, , \vartheta, 0)
\, \sim \, \frac{1}{r_{\perp}^{\,D - 2 + \eta_{\perp}}}
\end{equation}
with 
\begin{equation} \label{G110b}
\eta_{\perp} = \lambda_0 = \sqrt{3} \, \, .
\end{equation}
We note that within the present mean-field theory the relation
$2 \eta_{\perp} = \eta + \eta_{\parallel}$ is fulfilled 
because $\eta(D=4) = 0$. 

The exponents in Eqs.\,(\ref{G90}) and 
(\ref{G110}) for a thin needle are different from those for
the half-space for which $\eta_{\parallel}^{\,\text{hs}} = 6$
and $\eta_{\perp}^{\,\text{hs}} = 3$ in mean-field approximation.
Thus the correlations in the critical fluid are more suppressed 
in proximity of the needle than in the bulk fluid, and are
even more suppressed near the surface bounding 
the half-space. This can
be understood by noting that a needle represents a weaker, but 
still relevant, perturbation for the critical fluid than a planar 
wall (compare Fig.\,\ref{diagram}).  
For a sphere one finds that the two-point correlation 
function at criticality decays as
$| {\bf r} - {\bf r}\,' |^{- (D - 2 + \eta)}$ 
with the bulk exponent $\eta$ if 
the distance vector ${\bf r} - {\bf r}\,'$ is increased in any
direction. This reflects the fact that a sphere represents an 
irrelevant perturbation.

\end{mathletters}


\section{Amplitudes $\omega_{\pm}$ and 
scaling functions ${\cal F}_{\pm}(\lowercase{x}_{\pm})$}
\label{app_neu}

In this appendix we determine the universal amplitudes
$\omega_{\pm}$ (see Eqs.\,(\ref{E150}) and (\ref{E160}))
and the universal scaling functions
${\cal F}_{\pm}(x_{\pm})$
(see Eqs.\,(\ref{A240}) and (\ref{A250})).
The corresponding mean-field results
(see Eqs.\,(\ref{Q30}) and (\ref{M180}), respectively)
hold in $D = 4$.
For $D = 3$ and $2$ results are available 
\cite{MEF68,MEF73,TF75,TC75,bray76,liumef89}
for the Fourier transform of the bulk two-point correlation 
function introduced in Eq.\,(\ref{A250}), i.e.,
\begin{equation} \label{N10}
S_{t}(p) \, = \, \int d^D r \, e^{i \, {\bf p} \, {\bf r}} \, 
\langle \Phi({\bf r}) \Phi(0) \rangle_{b, \, t}^{\,C} \, \, \, ,
\end{equation}
which can be written in the scaling form
\begin{equation} \label{N20}
S_{t}(p) \, = \, C^{\pm} \, |t|^{-\gamma} \, 
g_{\pm}(p \, \xi_{\pm}) \, \, .
\end{equation}
The universal scaling functions $g_{\pm}(Y_{\pm})$ are fixed 
by the normalization condition $g_{\pm}(0) = 1$ and by the 
choice of $\xi_{\pm}$ as the true correlation length
for $T \gtrless T_c$.
The amplitudes $C^{\pm}$ are nonuniversal and 
$\gamma = \nu (2 - \eta)$ is a universal bulk
critical exponent. Numerical values of the universal
bulk critical exponent $\eta$ are given by
$\eta(D=2) = 1/4$ and $\eta(D=3) = 0.031$ 
\cite{zinn} (compare Eq.\,(\ref{A45})).
Inserting Eq.\,(\ref{A250}) into Eq.\,(\ref{N10}) and using
Eq.\,(\ref{E160}) yields
\begin{equation} \label{P10}
\omega_{\pm} \, = \, 
\frac{c_{\pm\,} 2^{\beta/\nu}}{\Omega_{D\,} \widetilde{Q}_{\pm}} 
\end{equation}
with $\Omega_{D} = 2 \pi^{D/2}/\,\Gamma(D/2)$. The universal
numbers $\widetilde{Q}_{\pm}$ are given by
\begin{equation} \label{P20}
\widetilde{Q}_{\pm} \, = \, \left\{\begin{array}{l@{\quad}l}
N \, R_{+}^{\,2-\eta} \, Q_3 \, \, \, , & T > T_c \, \, , \\[4mm]
N \, R_{+}^{\,2-\eta} \, R_{\xi}^{\,\eta - 2} \,
\displaystyle{\frac{C^{+}}{C^{-}}} \, Q_3 \, \, \, , & T < T_c \, \, ,
\end{array} \right.
\end{equation}
with the universal amplitude ratio
$Q_3 = \hat{D} \, (\xi_{0,1}^{\,+})^{2-\eta} / C^{+}$ 
introduced by Tarko and Fisher \cite{TF75}.
The amplitudes 
$\xi_{0,1}^{\,\pm}$ correspond to correlation length defined
via the second moment \cite{TF75} of the correlation function, 
i.e., $\xi_{\pm, 1} = \xi_{0,1}^{\,\pm\,} |t|^{-\nu}$. The
amplitude $\hat{D}$ is defined by 
$S_{t = 0}(p) = \hat{D} \, p^{\eta-2}$
which implies that $B_{\Phi}$ in Eq.\,(\ref{A200}) is related
to $\hat{D}$ via $B_{\Phi} = N \hat{D}$ with the numbers
\begin{equation} \label{P30}
N \, = \, \left\{ \begin{array}{l@{\quad}l}
\Gamma(\frac{1}{8}) / [ 2 \pi \, 2^{3/4\,} \Gamma(\frac{7}{8}) ] \, = \,
0.654308 \, \, \, , & D = 2 \, \, , \\[4mm]
\Gamma(\eta) \sin(\frac{\eta \, \pi}{2}) / (2 \pi^2) \, = \, 
0.078196 \, \, \, , & D = 3 \, \, .
\end{array} \right.
\end{equation}
Numerical values of $Q_3$ are given by $Q_3(D=2) = 0.41377$ 
and $Q_3(D=3) = 0.896$ \cite{TF75}.
In Eq.\,(\ref{P20}) $R_+ = \xi_0^{+} / \xi_{0,1}^{\,+}$, 
$R_{\xi} = \xi_0^{+} / \xi_0^{-}$, and $C^{+} / C^{-}$ are
known universal amplitude ratios with
$R_+(D=2) = 1.000402$ (see Eq.\,(3.7) in Ref.\,\cite{TC75}),
$R_+(D=3) = 1.0003$ \cite{MEF73};
$R_{\xi}(2) = 2$, $R_{\xi}(3) \simeq 1.92$ \cite{MEF73};
$(C^{+} / C^{-})(2) = 37.693562$ \cite{TF75}, 
$(C^{+} / C^{-})(3) = 4.95$ \cite{liumef89}. 
Using Eqs.\,(\ref{P10}) and (\ref{P20}) in conjunction with
the values of $c_{\pm}(D=2)$ from Table \ref{tab_ca} 
and with $c_+(D = 3) = 0.94$, and $c_{-}(D=3) = 1.24$ 
yields the values of $\omega_{\pm}$ quoted in Table \ref{tab_neu}.
We note that the accuracy of the quoted value of
$\omega_{-}(D=3)$ is unknown because the 
accuracy of the aforementioned value of $R_{\xi}(D=3)$ 
is not given reliably \cite{MEF73}.

Next we outline how the universal scaling functions
${\cal F}_{\pm}(x_{\pm})$ introduced in Eq.\,(\ref{A250})
can be inferred from presently available results for 
$g_{\pm}(Y_{\pm})$ (see Eqs.\,(\ref{N10}) and (\ref{N20})). 

(a) $D = 3$: In this case the scaling functions 
${\cal F}_{\pm}(x_{\pm})$ are given by
\begin{equation} \label{N30}
{\cal F}_{\pm}(x_{\pm}) \, = \, 
k_{\pm} \, x_{\pm}^{\,\eta}  \, 
\int\limits_{0}^{\infty} \, dY_{\pm} \, Y_{\pm} \, 
\sin(Y_{\pm} x_{\pm}) \, g_{\pm}(Y_{\pm}) \, \, , \quad
D = 3 \, \, ,
\end{equation}
with the amplitudes $k_{\pm}$ fixed by the condition
${\cal F}_{\pm}(0) = 1$ which allows one to express 
the nonuniversal amplitudes $C^{\pm}$ in terms of 
$B_{\Phi}$ and $\xi_{0}^{\pm}$.
For the case $T > T_c$ the approximation for $g_{+}(Y_+)$ 
proposed by Bray \cite{bray76} can be regarded to be reliable.
Accordingly, for values $Y_+ \le 20$ the function
$g_{+}(Y_+)$ can be inferred 
from the first column in Table V of
Ref.\,\cite{bray76} whereas for $Y_+ > 20$ the asymptotic
expansion by Fisher and Langer \cite{MEF68}
is applicable, i.e.,
\begin{equation} \label{N40}
g_{+}(Y_{+} \to \infty) \, = \, \frac{C_1}{Y_{+}^{\,2 - \eta}} 
\Big[ 1 + \frac{C_2}{Y_{+}^{\,(1 - \alpha)/\nu}}
+ \frac{C_3}{Y_{+}^{\,1/\nu}} + \, \ldots \, \Big] \, \, , \quad
D = 3 \, \, ,
\end{equation}
with coefficients $C_i$ and the bulk critical exponents
$\eta$, $\alpha$, and $\nu$.
Using in Eqs.\,(\ref{N30}) and (\ref{N40}) the 
values $\eta = 0.041$, $\nu = 0.638$, and $\alpha = 0.086$
as quoted in Ref.\,\cite{bray76} yields
$C_1 = 0.909$, $C_2 = 3.593$, $C_3 = - 4.493$ \cite{bray76}, 
and $k_{+} = 0.7166$. Unfortunately, for $T < T_c$ and $D = 3$ 
we are not aware of an accurate estimate of $g_{-}(Y_{-})$.

(b) $D = 2$: In this case the scaling functions 
${\cal F}_{\pm}(x_{\pm})$ are given by
\begin{equation} \label{N50}
{\cal F}_{\pm}(x_{\pm}) \, = \, 
k_{\pm}^{\,'} \, x_{\pm}^{\,\eta}  \, 
\int\limits_{0}^{\infty} \, dY_{\pm} \, Y_{\pm} \, 
J_0(Y_{\pm} x_{\pm}) \, g_{\pm}(Y_{\pm}) \, \, , \quad
D = 2 \, \, ,
\end{equation}
where $J_0$ is a modified Bessel function \cite{abr}. In $D = 2$
exact results for $g_{\pm}(Y_{\pm})$ can be deduced 
from Ref.\,\cite{TC75}. However, some care is necessary regarding
the definition of the scaling variable $y$ used in Ref.\,\cite{TC75}:
whereas $y = Y_{+}$ for $T > T_c$, one has 
$y = 1.959 \, Y_{-}$ for $T < T_c$.
The numerical prefactor in front of $Y_{-}$ equals
$(\sqrt{\Sigma_2^{-}} \, R_{-})^{-1}$
with $\Sigma_2^{-}$ from Table IV in Ref.\,\cite{TC75}
(see Eq.\,(3.7) in Ref.\,\cite{TC75})
and with the universal amplitude ratio 
$R_{-} = \xi_0^{-} / \xi_{0,1}^{\,-} = 1.615$ in $D = 2$ 
\cite{MEF73}. For the amplitudes $k_{\pm}^{\,'}$ in 
Eq.\,(\ref{N50}) one obtains $k_{+}^{\,'} = 0.5874$ and 
$k_{-}^{\,'} = 0.05055$.


\section{Small curvature expansion of the excess adsorption} 
\label{appB}

In this appendix we outline
how the coefficients $a_{d,D}^{\,\pm}$ and 
$b_{d,D}^{\,\pm}$ introduced in Eq.\,(\ref{E110}) can be calculated 
within mean-field theory, i.e., for $D = 4$. Since the approaches
for $T > T_c$ ($+$) and for $T < T_c$ ($-$) are quite similar we 
restrict our presentation to the case $T > T_c$. 

According to Eqs.\,(\ref{M40}) and (\ref{M80}) for $D = 4$
the universal scaling function $P_+(x_+, y_+)$ satisfies
the nonlinear differential equation (in the following we drop the 
subscript `$+$' in order to simplify the notation)
\begin{mathletters} \label{B10}
\begin{equation} \label{B10a}
P''(x,y) \, + \, \frac{d-1}{x+y} \, P'(x,y) \, - \, P(x,y) 
\, = \, P^{3}(x,y)
\end{equation}
where the derivatives are taken with respect to the variable $x$.
The parameter $d$ with $1 \le d \le 4$ can be chosen
arbitrarily. Equation (\ref{B10a}) is supplemented 
by the boundary conditions 
\begin{equation} \label{B10b}
P(x \to 0, y) \, \to \, \frac{\sqrt{2}}{x} \, \, , \qquad
P(\infty, y) \, = P_{+,\,b} \, = \, 0 \, \, .
\end{equation}
The scaling function $G(y) \equiv G_+(y_+)$ from Eq.\,(\ref{E60})
then reads
\end{mathletters}
\begin{equation} \label{B20}
G(y) \, = \, 
\int\limits_{0}^{\infty} dx \,
\Big\{ \Big( \frac{x}{y} + 1 \Big)^{d - 1}
P(x,y) \, - \, P_0(x) \Big\}
\end{equation}
with $P(x,y)$ defined by Eq.\,(\ref{B10}) and the 
half-space profile $P_0(x) \equiv P(x,\infty) = P_+(x_+)$ 
given by the first part of Eq.\,(\ref{A30}).
Note that the singularity at $x = 0$ of the first term in curly brackets 
in Eq.\,(\ref{B20}) is cancelled by the second term. According
to Eq.\,(\ref{E110}) the function $G(y)$ can be expanded as
\begin{equation} \label{B30}
G(y \to \infty) \, = \, 
a_d \, y^{-1} \, + \, b_d \, y^{-2} \, + \, \ldots
\end{equation}
with the coefficients 
$a_d \equiv a_{d,4}^{\,+}$ and $b_d \equiv b_{d,4}^{\,+}$
which we want to determine. 
To this end we expand $P(x,y)$ as
\begin{equation} \label{B40}
P(x, y \to \infty) \, = \, 
P_0(x) \, + \, P_1(x) \, y^{-1} \, + \, P_2(x) \, y^{-2} 
\, + \, \ldots \, \, .
\end{equation}
By inserting Eq.\,(\ref{B40}) into Eq.\,(\ref{B10a}) and 
equating terms with the same power in $y$ one derives the 
familiar nonlinear differential equation 
for the half-space profile $P_0(x)$, i.e., 
\begin{mathletters} \label{B50}
\begin{equation} \label{B50a}
P_0'' \, - \, P_0 \, = \, P_0^{\,3} \, \, ,
\end{equation}
\begin{equation} \label{B50b}
P_0(x \to 0) \, \to \, \frac{\sqrt{2}}{x} \, \, , \qquad
P_0(\infty) \, = \, 0 \, \, ,
\end{equation}
and the two linear differential equations 
\end{mathletters}
\begin{mathletters} \label{B60}
\begin{equation} \label{B60a}
P_1'' \, - \, P_1 \, - \, 3 P_0^{\,2} P_1
\, = \, - (d-1) P_0' \, \, ,
\end{equation}
\begin{equation} \label{B60b}
P_1(0) \, = \, - \frac{\sqrt{2}}{6} \, (d-1) 
\, \, , \qquad P_1(\infty) \, = \, 0 \, \, ,
\end{equation}
and 
\end{mathletters}
\begin{mathletters} \label{B70}
\begin{equation} \label{B70a}
P_2'' \, - \, P_2 \, - \, 3 P_0^{\,2} P_2 \, = \,
- (d-1) P_1' \, + \, 3 P_0 P_1^{\,2} \, + \, 
(d - 1) \, x P_0' \, \, ,
\end{equation}
\begin{equation} \label{B70b}
P_2'(0) \, = \, - \, \frac{\sqrt{2}}{36} \, (d-1)^2 \, + \,
\frac{\sqrt{2}}{6} \, (d-1) \, \, , \qquad
P_2(\infty) \, = \, 0 \, \, ,
\end{equation}
for $P_1(x)$ and $P_2(x)$, respectively. For the boundary
conditions in the second parts of the above equations compare
Eqs.\,(\ref{M150}) and (\ref{M80}). By using
\end{mathletters}
\begin{equation} \label{B80}
\Big( \frac{x}{y} + 1 \Big)^{d - 1} \, = \,
1 \, + \, (d - 1) \, x \, y^{-1} \, + \,
\frac{1}{2} \, (d - 1) (d - 2) \, x^2 \, y^{-2} \, + \, \ldots
\end{equation}
in the integrand on the rhs of Eq.\,(\ref{B20})
in conjunction with Eq.\,(\ref{B40}) we find for the 
coefficients in Eq.\,(\ref{B30}) the expressions 
\begin{equation} \label{B90}
a_d \, = \, (d-1) \int\limits_{0}^{\infty} dx \, x \, P_0(x)
\, + \, \int\limits_{0}^{\infty} dx \, P_1(x)
\end{equation}
and
\begin{eqnarray}
& & b_d \, = \, \frac{1}{2} \, (d-1)(d-2) 
\int\limits_{0}^{\infty} dx \, x^2 \, P_0(x) \label{B100}\\
& & \, + \,
(d - 1) \int\limits_{0}^{\infty} dx \, x \, P_1(x) \, + \,
\int\limits_{0}^{\infty} dx \, P_2(x) \nonumber
\end{eqnarray}
where $P_0(x)$, $P_1(x)$, and $P_2(x)$ are the solutions of
Eqs.\,(\ref{B50}) - (\ref{B70}). While $P_0(x)$ is given
by the first part of Eq.\,(\ref{A30}) we now
turn to the calculation of 
$P_1(x)$ and $P_2(x)$. It is important
to retain the dependence of the latter functions on $d$
in analytical form in order to 
be able to carry out 
the consistency check of Eq.\,(\ref{E125})
for the present case $D = 4$.

(i) Function $P_1(x)$. 
Both the rhs of Eq.\,(\ref{B60a}) and the
boundary condition in the first part of Eq.\,(\ref{B60b})
depend on $d$ via the term $(d-1)$. We
note that even the full function 
$P_1(x)$ exhibits this simple dependence on $d$. 
The reason is that Eq.\,(\ref{B60a}) is linear with respect 
to $P_1$ so that Eq.\,(\ref{B60}) is solved by the ansatz
\begin{equation} \label{B110}
P_1(x) \, = (d - 1) \, U(x)
\end{equation}
where $U(x)$ is independent of $d$ and satisfies
the inhomogeneous linear differential equation
\begin{mathletters} \label{B120}
\begin{equation} \label{B120a}
U'' \, - \, U \, - \, 3 P_0^{\,2} U
\, = \, - P_0' \, \, ,
\end{equation}
\begin{equation} \label{B120b}
U(0) \, = \, - \frac{\sqrt{2}}{6} 
\, \, , \qquad U(\infty) \, = \, 0 \, \, .
\end{equation}
The solution of Eq.\,(\ref{B120}) can be 
given analytically (see, e.g., Ref.\,\cite{bender}):
\end{mathletters}
\begin{eqnarray}
U(x) \, & = & \, 
\frac{\cosh(x) - \sinh(x)}{6 \sqrt{2} \sinh^2(x)} \label{B130}\\
& & \times \, \Big[ 2 \cosh(2 x) - 2 - 3 x - 3 x \cosh(2 x) + 
3 \sinh(2 x) - 3 x \sinh(2 x) \Big] \, \, . \nonumber
\end{eqnarray}

(ii) Function $P_2(x)$. By using Eq.\,(\ref{B110}) the 
differential equation (\ref{B70a}) turns into 
\begin{equation} \label{B140}
P_2'' \, - \, P_2 \, - \, 3 P_0^{\,2} P_2 \, = \,
- (d-1)^2 \, U' \, + \, (d-1)^2 \, 3 P_0 U^2 \, + \,
(d - 1) \, x P_0' 
\end{equation}
with $U(x)$ from Eq.\,(\ref{B130}). In this case
both on the rhs of Eq.\,(\ref{B140}) and in the
boundary condition in the first part of Eq.\,(\ref{B70b})
different dependences on $d$ arise via the
terms $(d-1)^2$ and $(d-1)$. Again it is crucial to observe
that Eq.\,(\ref{B140}) is linear with respect to $P_2$ so 
that its dependence on $d$ takes the simple form
\begin{equation} \label{B150}
P_2(x) \, = \, (d-1)^2 \, V(x) \, + \, (d-1) \, W(x)
\end{equation}
where $V(x)$ and $W(x)$ are independent of $d$ and satisfy 
separately the inhomogeneous linear differential equations
\begin{mathletters} \label{B160}
\begin{equation} \label{B160a}
V'' \, - \, V \, - \, 3 P_0^{\,2} V \, = \,
- U' \, + \, 3 P_0 U^2 
\end{equation}
\begin{equation} \label{B160b}
V'(0) \, = \, - \, \frac{\sqrt{2}}{36} \, \, , \qquad
V(\infty) \, = \, 0 \, \, ,
\end{equation}
and 
\end{mathletters}
\begin{mathletters} \label{B170}
\begin{equation} \label{B170a}
W'' \, - \, W \, - \, 3 P_0^{\,2} W \, = \, x P_0'
\end{equation}
\begin{equation} \label{B170b}
W'(0) \, = \, \frac{\sqrt{2}}{6} \, \, , \qquad
W(\infty) \, = \, 0 \, \, .
\end{equation}
Equations (\ref{B160}) and 
(\ref{B170}) can be solved numerically, which is 
facilitated by the fact that both $U(x)$ and $P_0(x)$
on the rhs of Eqs.\,(\ref{B160a}) and (\ref{B170a}) are known
in analytical form. By inserting the resulting function $P_2(x)$
from Eq.\,(\ref{B150}) and the function $P_1(x)$ from Eq.\,(\ref{B110})
into Eqs.\,(\ref{B90}) and (\ref{B100}) one obtains the 
dependences of $a_{d,4}^{\,+} = a_d$ and $b_{d,4}^{\,+} = b_d$ 
on $d$ as given by Eq.\,(\ref{E125}) with the curvature parameters
given in the first parts of Eq.\,(\ref{Q20}). The procedure 
in the case $T < T_c$ is completely analogous to that 
for $T > T_c$ as outlined above and yields
the curvature parameters in the second parts of
Eq.\,(\ref{Q20}).

\end{mathletters}



\begin{thebibliography}{99}
\bibitem{definition}Any definition $\xi_{\pm}^{\,\,'}$ different from that
of $\xi_{\pm}$ as the true correlation length is related to $\xi_{\pm}$
by the universal ratio $\xi_{\pm}^{\,\,'} / \xi_{\pm}$. This leads to 
a corresponding change of the scaling functions $P_{\pm}$ 
and of quantities obtained from them 
so that experimentally observable quantities 
remain unchanged (for more details see 
Appendix \ref{app_neu} and Ref.\,\cite{floter}).
\bibitem{bar}The bars on $\bar{a}_{\pm}$, $\bar{a}_{\pm}^{\,'}$, 
$\bar{b}_{\pm}$ serve to distinguish these coefficients from 
the corresponding ones
$a_{\pm}$, $a_{\pm}^{\,'}$, $b_{\pm}$ introduced in Ref.\,\cite{ds93}
which are defined differently from Eq.\,(\ref{A40}).
\bibitem{coth}The curves for $P_{-}(x_{-})$ for $D = 4$
shown in Fig.\,2 of Ref.\,\cite{sdl94} and in Fig.\,1\,(b) 
of Ref.\,\cite{floter}, which do not comply with this 
statement, are not correct.
\bibitem{geometry}We briefly outline the argument: Since the 
curvature contributions must depend on how the surface is embedded 
in the space ${\mathbb R}^{D}$ they must be derivable from the local 
extrinsic curvature tensor ${\bf K}_{ij} = K_{ij\,} {\bf n}$
where ${\bf n}$ is the local unit vector normal to the surface 
(we only consider orientable surfaces). The principal local radii of
curvature $R_i$ are the inverse of the $D - 1$ eigenvalues of the
matrix $(K_{ij})$. Therefore $K_m$, $K_m^{\,2}$,
and $K_G$ as defined in Eq.\,(\ref{A100}) are the only independent 
scalar quantities to first and second order in $1 / R_i$ 
which can be deduced from $K_{ij}$ and which are invariant under
permutations of the indices $i = 1, \, \ldots \, , D - 1$.
\bibitem{rapid}The leading regular term captured by the 
ellipses in Eq.\,(\ref{A150}) which is analytic in $1/R$
is proportional to $t / R \sim \xi_{\pm}^{\,- 1 / \nu} / R$. 
Since all terms in curly brackets depend only on the 
dimensionless variables $x_{\pm}$ and $y_{\pm}$ this regular 
term must scale with $x_{\pm}^{\,1 + \,1/\nu}$ where the 
exponent is larger than two if $D > 2$.
\bibitem{shape}The dashed line in 
Fig.\,\ref{diagram} is known rather accurately by means
of the $\varepsilon$-expansions of $\beta(D)$ and $\nu(D)$
in conjunction with the corresponding exact values 
in $D = 2$ \cite{zinn}.
\bibitem{line}In a lattice model 
such a `cylinder' with microscopically small
radius $R$ can be realized by 
a single line of lattice sites where the orientation 
of the spins is fixed.
\bibitem{shoot}It is instructive to interpret Eq.\,(\ref{M40})
as an equation of motion of a particle at `position' $m$ and 
`time' $s$ subjected to the `potential energy' 
$- \tau m^2 / 2 - m^4 / 4$ \cite{binder,diehl}. In general
the differential equation cannot be solved analytically due 
to the time-dependent friction term. One seeks 
solutions $m(s)$ which come at rest on the top $m_b$ of the 
potential hill at time $s \to \infty$. This problem
can be solved 
numerically, e.g., by means of a `shooting method'
as a trial and error approach.
\bibitem{interior}The curvature parameters for a convex 
surface as considered here should, of course, be the same
as for a concave surface. In the present mean-field 
case this does indeed hold. This can be confirmed by using
the methods outlined in Appendix \ref{appB}.
\end{thebibliography}
\end{document}